\documentclass[aps,
               pra,
               twocolumn,
               amsmath,
               amssymb,
               superscriptaddress,
               reprint,
               10pt,
               nofootinbib
              ]{revtex4-2}

\usepackage[hyperindex,colorlinks,breaklinks,allcolors=blue]{hyperref}
\usepackage{txfonts}

\usepackage{amsmath, amssymb}
\usepackage{pstricks}
\usepackage{graphicx}
\usepackage{tikz}
\usepackage[utf8x]{inputenc}
\usepackage{color}
\usepackage{bm}

\usepackage{pstool}
\usepackage{pgf}
\usepackage{import}

\usepackage{cleveref}

\usetikzlibrary{snakes}
\usetikzlibrary{arrows}

\usepackage{tabularx}
\usepackage{multirow}
\usepackage{array}
\newcolumntype{L}[1]{>{\raggedright\let\newline\\\arraybackslash\hspace{0pt}}m{#1}}
\newcolumntype{C}[1]{>{\centering\let\newline\\\arraybackslash\hspace{0pt}}m{#1}}
\newcolumntype{R}[1]{>{\raggedleft\let\newline\\\arraybackslash\hspace{0pt}}m{#1}}

\newcommand\T{\rule{0pt}{2.6ex}}       
\newcommand\B{\rule[-1.2ex]{0pt}{0pt}} 

\newcommand{\eq}[1]{(\ref{eq:#1})}
\newcommand{\Eq}[1]{Eq.\,\eqref{eq:#1}}

\newcommand{\Fig}[1]{Fig.~\ref{fig:#1}}

\newcommand{\app}[1]{\ref{app:#1}}
\newcommand{\App}[1]{App.~\ref{app:#1}}
\newcommand{\Tab}[1]{Table~\ref{tab:#1}}
\newcommand{\eg}{\textit{e.\,g.}}
\newcommand{\zh}{z_\text{h}}

\definecolor{ao(english)}{rgb}{0.0, 0.5, 0.0}
\definecolor{applegreen}{rgb}{0.55, 0.71, 0.0}
\definecolor{cadetblue}{rgb}{0.37, 0.62, 0.63}
\definecolor{cadet}{rgb}{0.33, 0.41, 0.47}
\definecolor{byzantine}{rgb}{0.74, 0.2, 0.64}
\definecolor{orange}{rgb}{1.0, 0.5, 0.0}

\newcommand{\ie}{\textit{i.\,e.}}
\newcommand{\cf}{\textit{cf.}}

\renewcommand{\i}{\mathrm{i}}

\newcommand{\footnoteremember}[2]{%
\footnote{#2}%
\newcounter{#1}%
\setcounter{#1}{\value{footnote}}%
}

\newcommand{\zc}{\mathfrak{z}}

\newcommand{\lengthscale}{\ell'}
\newcommand{\gridlengthscale}{\ell}
\newcommand{\dimless}[1]{{#1}'}
\newcommand{\gridunit}[1]{\bar{#1}}

\makeatletter
\let\cat@comma@active\@empty
\makeatother

\begin{document}

\title{Vortex motion quantifies strong dissipation in a holographic superfluid}

\author{Paul Wittmer}
\thanks{These two authors contributed equally.}
\affiliation{Institut f\"ur Theoretische Physik, 
              Universit\"at Heidelberg,
              Philosophenweg 16, 
             69120 Heidelberg, Germany}
\affiliation{ExtreMe Matter Institute EMMI, 
             GSI Helmholtzzentrum  f\"ur Schwerionenforschung,
             Planckstrasse 1, 
             64291 Darmstadt, Germany}
\author{Christian-Marcel Schmied}
\thanks{These two authors contributed equally.}
\affiliation{ExtreMe Matter Institute EMMI, 
	GSI Helmholtzzentrum  f\"ur Schwerionenforschung,
	Planckstrasse 1, 
	64291 Darmstadt, Germany}   
\affiliation{Kirchhoff-Institut f\"ur Physik, 
	Universit\"at Heidelberg,
	Im Neuenheimer Feld 227, 
	69120 Heidelberg, Germany}
\author{Thomas Gasenzer}
\affiliation{Institut f\"ur Theoretische Physik, 
             Universit\"at Heidelberg,
             Philosophenweg 16, 
             69120 Heidelberg, Germany}
\affiliation{ExtreMe Matter Institute EMMI, 
	GSI Helmholtzzentrum  f\"ur Schwerionenforschung,
	Planckstrasse 1, 
	64291 Darmstadt, Germany}   
\affiliation{Kirchhoff-Institut f\"ur Physik, 
	Universit\"at Heidelberg,
	Im Neuenheimer Feld 227, 
	69120 Heidelberg, Germany}
\author{Carlo Ewerz}	
\affiliation{Institut f\"ur Theoretische Physik, 
             Universit\"at Heidelberg,
             Philosophenweg 16, 
             69120 Heidelberg, Germany}
\affiliation{ExtreMe Matter Institute EMMI, 
             GSI Helmholtzzentrum  f\"ur Schwerionenforschung,
             Planckstrasse 1, 
             64291 Darmstadt, Germany}

\begin{abstract}
Holographic duality provides a description of strongly coupled quantum systems in terms of weakly coupled gravitational theories in a higher-dimensional space. 
It is a challenge, however, to quantitatively determine the physical parameters of the quantum systems corresponding to generic holographic theories. 
Here, we address this problem for the two-dimensional holographic superfluid, known to exhibit strong dissipation. We numerically simulate the motion of a vortex dipole and perform a high-precision matching of the corresponding dynamics resulting from the dissipative Gross-Pitaevskii equation. 
Excellent agreement is found for the vortex core shape and the spatio-temporal trajectories. 
A further comparison to the Hall-Vinen-Iordanskii equations for point vortices interacting with the superfluid allows us to determine the friction parameters of the holographic superfluid. 
Our results suggest that holographic vortex dynamics can be applied to experimentally accessible superfluids like strongly coupled ultracold Bose gases or thin helium films with temperatures in the Kelvin range. 
This would make holographic far-from-equilibrium dynamics and turbulence amenable to experimental tests. 
\end{abstract}

\pacs{%
  03.75.Lm,       
11.25.Tq,	
67.40.Vs	
}

\maketitle


\textit{Introduction}. 
The time evolution of quantum many-body systems out of equilibrium
has attracted considerable attention in recent years
\cite{Sachkou2019a.Science366.1480,Johnstone2019a.Science.364.1267,Gauthier2019a.Science.364.1264,Prufer:2018hto,Eigen2018a.arXiv180509802E,Erne:2018gmz,Harris2016a.NatPhys8.788,Chomaz2015a.NatComm.6.6162}.
Strong correlations prevailing in these systems generically necessitate nonperturbative methods to quantitatively describe the dynamics.
In particular the interplay between linear and strongly nonlinear excitations such as topological defects
poses a challenge for theory which is even amplified for strongly coupled and dissipative systems. 

Holography \cite{Maldacena1999,Gubser1998a,Witten:1998qj}, also known as gauge-gravity duality, allows one to address such problems in an intrinsically nonperturbative framework. It posits the equivalence of certain quantum field theories to gravitational theories with an additional dimension of space. In this duality, strongly coupled field theory is mapped to weakly coupled, and hence classical, Einstein gravity on an Anti-de Sitter (AdS) space. Finite temperature of the field theory corresponds to a black hole on the gravity side. By now, a wide range of holographic dualities has been established, with applications ranging from nuclear to condensed matter physics. While the systems described by generic holographic models are qualitatively known to be strongly coupled, it has proven notoriously difficult to quantitatively determine their phenomenological parameters. 

Here, we address this longstanding problem for the holographic model of a superfluid in two spatial dimensions \cite{Gubser:2008px, Hartnoll:2008vx, Herzog:2008he,Hartnoll:2016apf}. This system has been studied extensively, with particular focus on linear \cite{Hartnoll:2008vx,Herzog:2008he,Sonner2010a.PhysRevD.82.026001,Anninos:2010sq} and nonlinear excitations such as vortices \cite{Bhaseen:2012gg,Keranen:2009re,Dias:2013bwa,Chesler2013a.Science341.368,Ewerz:2014tua,Lan:2018llf,Ewerz:2020}. So far it was unknown how the holographic superfluid compares to experimentally realized superfluids like ultracold atomic gases \cite{Pitaevskii2003a} or liquid helium \cite{Donnelly2005a}. 

Specifically, we aim at a quantitative characterization of vortex dynamics in the holographic superfluid. 
For that we study the time-evolution of a vortex-antivortex pair in holography and in the dissipative Gross-Pitaevskii equation (DGPE) \cite{Proukakis_2008}. 
By tuning the DGPE parameters, we match the holographic vortices' core shape as well as their trajectories in space and time \footnote{For videos of the vortex dynamics see \href{https://www.thphys.uni-heidelberg.de/~holography/holoDGPE/}{https://www.thphys.uni-heidelberg.de/$\sim$holography/holoDGPE/}.}.
We furthermore use known relations between the DGPE and the Hall-Vinen-Iordanskii (HVI) equations \cite{Hall1956a.PRSLA.238.204,Iordansky1964a.AnnPhys.29.335,Iordanskii1966JETP...22..160I} describing the mechanical motion of point vortices subject to interactions with the superfluid. 
This allows extracting friction coefficients of the holographic superfluid which we compare to those of experimentally accessible superfluids. 
In the following we present our main results. An appendix describes all technical details. 

\textit{Superfluidity} is a low-temperature phenomenon
associated with Bose-Einstein condensation. 
The condensate is described by a non-zero expectation value of a bosonic field operator, $\langle\Psi(\mathbf{r}, t)\rangle\not=0$. 
The classical field $\psi(\mathbf{r},t)=\langle\Psi(\mathbf{r},t)\rangle = \sqrt{\rho(\mathbf{r},t)} \exp\{\i \theta(\mathbf{r},t)\}$ 
encodes the density $\rho = \lvert\psi\rvert^2$ of the condensed particles and their velocity field $\mathbf{v}(\mathbf{r},t)=\nabla\theta(\mathbf{r},t)$.
$\psi$ acts as an order parameter for the superfluid phase.
The Tisza-Landau two-fluid model \cite{Tisza1938TPiHII,PhysRev.60.356} offers a successful description of superfluidity by invoking a second, `normal' (or thermal) component of the total liquid and accounting for its interaction with the superfluid condensate.
We denote the thermal equilibrium condensate density as $\rho_0$ and use it as a background for imprinting vortices.

\textit{In holography}, a
$(2+1)$-dimensional super\-fluid has a dual gravitational description in terms of an 
Abelian Higgs model, 
\begin{align}\label{eq:AdSCFTLagrangian}
\mathcal{L}_{\text{gauge-matter}} &= - \frac 1 4 F_{\mu \nu} F^{\mu \nu}- \lvert D_{\mu} \Phi \rvert^2 - m^2 \lvert \Phi \rvert^2\,,
\end{align}
on a $(3+1)$-dimensional (asymptotically) AdS spacetime with a black hole \cite{Gubser:2008px,Hartnoll:2008vx,Herzog:2008he}.
In the probe limit, valid at sufficiently high temperatures, the energy of the gauge-matter sector is small enough to neglect its backreaction on the AdS spacetime. 
The model \eqref{eq:AdSCFTLagrangian} is solved in 
a static Schwarzschild-AdS background 
with curvature radius $L_{\mathrm{AdS}}$ and horizon position $z_\text{h}$,
\begin{align}\label{eq:Metric2DEF}
\text{d} s^2=\frac{L_\text{AdS}^2}{z^2}\left[- \left(1-\frac{z^3}{z_\text{h}^3}\right)\text{d} t^2+\text{d}x^2+\text{d}y^2-2\text{d} t\,\text{d} z\right]\,.
\end{align}
$t, x, y$ are the superfluid's coordinates, $z$ is the additional holographic coordinate, and we write $\bm{r}=(x,y)$. 
$\mu, \nu = t, x, y, z$ specify the vector field's components. 
The matter part \eqref{eq:AdSCFTLagrangian} contains a scalar field $\Phi$ with mass $m$, the field strength tensor $F_{\mu \nu} = \nabla_{\mu} A_{\nu} -  \nabla_{\nu} A_{\mu}$ of the gauge field $A_{\mu}$, and 
the gauge-covariant derivative $D_{\mu}= \nabla_{\mu} - i A_{\mu}$.

\begin{figure}
	\includegraphics[width=0.7\columnwidth]{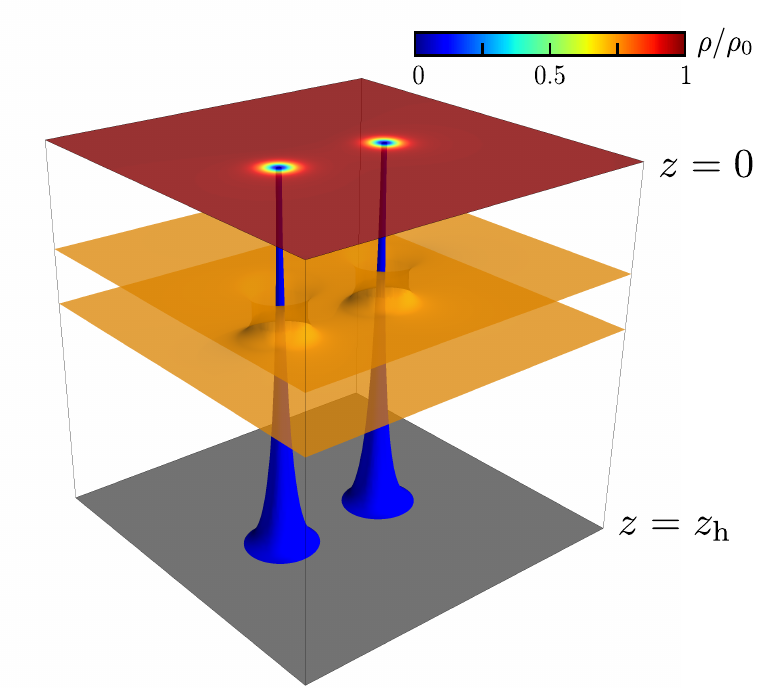}
	\caption{%
		Visualization of the bulk configuration of the 
		holographic superfluid in the presence of a 
		vortex-antivortex pair. 
		The blue tubes are isosurfaces of the scalar field $|\Phi|^2/z^4$, which reduces to the 
		superfluid density $\rho$
                at the boundary ($z=0$).
		The normalized superfluid density $\rho/\rho_0$ is encoded according to the color scale.
		Isosurfaces of the scalar charge cloud in the bulk are illustrated by the two orange sheets which the vortex tubes pierce through.
		The grey area at $z=z_\text{h}$ represents the black hole horizon.
		\label{fig:BulkGeometry}
	}
\end{figure}

Spontaneous symmetry breaking occurs in the holographic superfluid due to condensation of the scalar field near the black hole and formation of a charge cloud in the bulk of the AdS spacetime, screening the boundary from the black hole \cite{Gubser:2008px}. The superfluid can be thought of as a projection of the bulk dynamics onto the boundary of the spacetime at $z=0$, with the quantum expectation value $\psi=\langle\Psi\rangle$ 
obtained from a near-boundary expansion of the dual field $\Phi$ under appropriate boundary conditions,
$\Phi(t, \bm{r}, z)=
\psi(\bm{r},t) \,z^2 +\mathcal{O}(z^3)$. 
\Fig{BulkGeometry} illustrates the bulk picture. 
In the duality, the black hole corresponds to a static heat bath with temperature $\tilde{T} = 3/(4 \pi z_\text{h})$ and 
may be considered as the normal component of the system, whereas the gauge-matter sector corresponds to the superfluid component.
The chemical potential of the superfluid is fixed by the boundary condition of the temporal gauge field component, $\tilde{\mu}=A_t (z=0)$
\footnoteremember{holotilde}{To distinguish quantities denoted by the same letter, they are written with a tilde in holography and without in DGPE.}. 
Notably, $\tilde{T}$ and $\tilde{\mu}$ are not independent here. 
In our numerical simulations we choose $z_\text{h}\equiv1$ such that $\tilde{\mu}$ is the only free parameter, which sets the system into the superfluid phase above a critical value $\tilde\mu_\mathrm{c}\simeq4.064$ \cite{Herzog:2008he}.
It further fixes the ratio $\tilde{T}/\tilde{T}_\mathrm{c}$ via $\tilde{\mu}\tilde{T}/\tilde{T}_\mathrm{c}=\tilde\mu_\mathrm{c}$.

\textit{Dissipative Gross-Pitaevskii model}. 
A dilute cold Bose gas subject to damping due to interactions between the condensate and the non-condensate excitations can be described by the
DGPE for the order-parameter field $\psi(\mathbf{r},t)$. 
For a single-component system it reads
\begin{equation} \label{eq:DGPE}
\partial_t \psi(\mathbf{r},t) = - \left( {\i + \gamma} \right) \left [ - \frac {1}{2M }  {\nabla^2} + g\, \lvert \psi (\mathbf{r},t) \vert^2 - \mu \right] \psi(\mathbf{r},t)\,,
\end{equation}
where $M$ is the mass of the bosonic particles.
The coupling parameter $g$ characterizes the interactions of the particles, and $\mu$ is a chemical potential representing a constant shift of the single-particle energy.
The dimensionless phenomenological damping parameter $\gamma$ quantifies the dissipation.
We take $\mu = g \rho_0$ such that the zero mode is not damped. 
The healing length $\xi$, the characteristic length scale set by the interactions, is then given by $\xi = (2M \mu)^{-1/2}$.

\textit{Quantized vortices}
represent topological structures in the complex field $\psi$.
Around a vortex core, the phase of the order parameter winds by $2 \pi w_{i}$, with winding number $w_{i}\in\mathbb{Z} \backslash \{ 0 \}$.
Consequently, the density $\rho$ at the position of the core drops to zero, within a distance on the order of $\xi$. 
At large distances from the core,
$\rho$ approaches the background density $\rho_0$.

\textit{Numerical solution and matching procedure}. 
Both the holographic and DGPE systems are numerically solved on grids with $512 \times 512$ grid 
points in the $(x, y)$-direction subject to periodic boundary conditions.
For the holographic system, we additionally employ $32$ collocation points along the $z$-direction.
We consider several values of $\tilde\mu$ for which the probe limit is justified. 
For given $\tilde\mu$, we then adjust the numerical parameters in the DGPE simulations to
match the vortices' sizes and space-time trajectories from holography.  

To match the vortex sizes, we imprint a symmetric and periodic, and hence static, $2 \times 2$ configuration of two vortices and two antivortices with winding numbers $\pm 1$ into a homogeneous condensate background $\rho_0$. (This configuration is used only for this purpose.) 
We tune the DGPE healing length $\xi$ to adjust the size of the vortices to the
holographic ones.

Using $\xi$ determined from the vortex sizes, we then study the propagation of one vortex-antivortex pair which without 
friction would move perpendicular to the dipole vector $\mathbf{d}$ connecting the vortex positions.
In presence of friction between the defects and the fluid, the vortices are slowed relative to the superfluid flow such that a Magnus force emerges, causing a velocity component parallel to $\mathbf{d}$.
As a result, the vortices approach each other and annihilate into a rapidly decaying arc wave. 
We determine the trajectories on the grid with sub-plaquette resolution in a quasi-continuous manner by a combination of two tracking methods, \cf\ \cite{Ewerz:2020}: a two-dimensional Gaussian fit to the density depression around the vortex cores and a Newton-Raphson algorithm for tracking the zeros in the superfluid density.
The precision of this procedure allows us to match the DGPE vortex trajectories in space and time by adjusting the dissipation constant $\gamma$ and time rescaling parameter $\tau$. (The latter defines the DGPE grid unit
of time relative to the spatial grid unit $s$ by $t/\gridunit{t}=M s^{2}/\tau$.) 

With the knowledge of the DGPE parameters, we can fit solutions of the HVI equations for the motion of point vortices to the dipole trajectories to obtain friction coefficients of the holographic superfluid. We do this in a regime where effects due to the preparation of the initial vortex configuration (in particular a slight initial outward bending of the trajectories \cite{Ewerz:2020}) have died out but the vortices are still far enough apart.

\begin{figure}[t]
	\includegraphics[width=0.9\columnwidth]{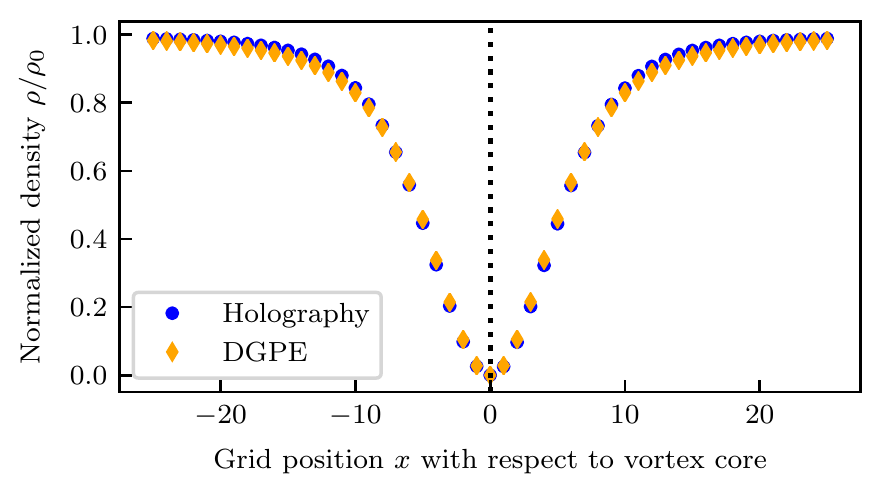}
	\caption{%
		Normalized density profile $\rho/\rho_0$ of a 
		single vortex on a one-dimensional cut through its center, in holography ($\tilde{\mu} = 6$) and DGPE. 
		The vortex shapes agree within $2 \%$ of the background density.
		\label{fig:VortexShape}
	}
\end{figure}

\textit{Results}.
Matching the sizes of the vortices, we find also their shapes in remarkable agreement, see \Fig{VortexShape}.
We furthermore obtain that the directly fitted width $\tilde\xi$ of the holographic 
vortex cores scales as $\tilde\xi\sim(\tilde\mu-\tilde\mu_{0})^{-1/2}$, with the 
shift $\tilde\mu_{0}\simeq4.06$ being close to the critical chemical 
potential $\tilde\mu_\mathrm{c}\simeq4.064$,
as is typical for nonrelativistic superfluids. 

\begin{figure}[t]
	\includegraphics[width=0.9\columnwidth]{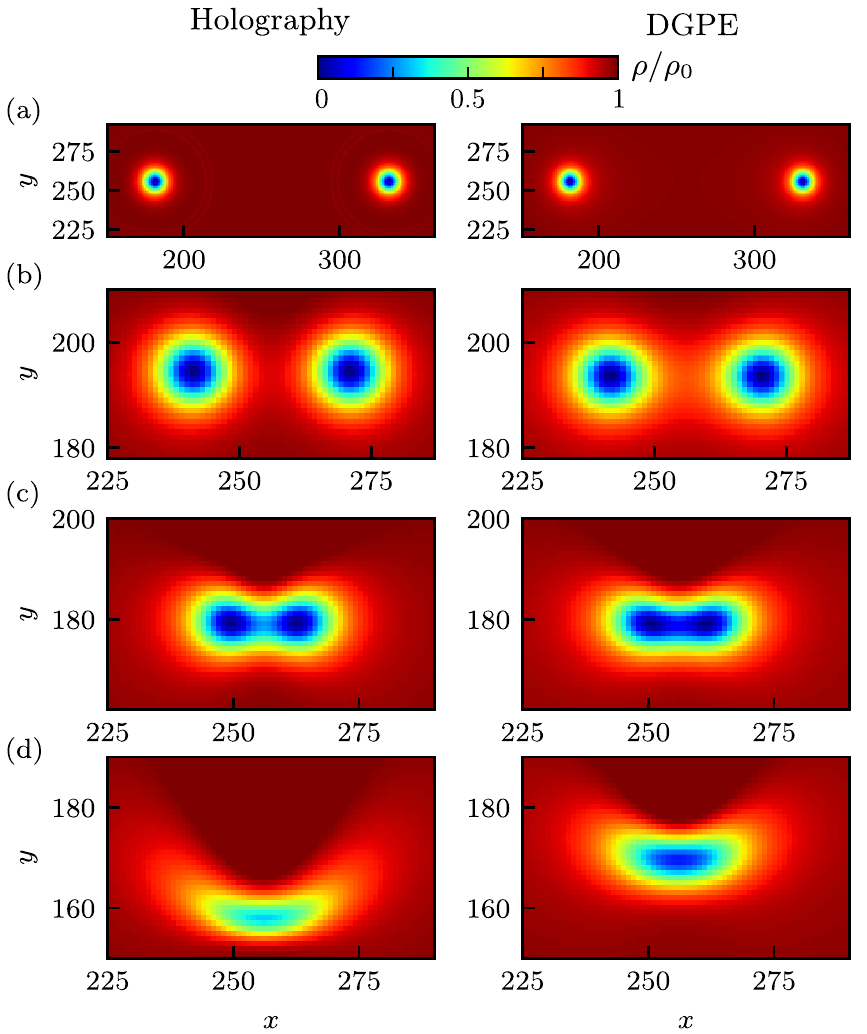}
	\caption{%
		Snapshots of the normalized superfluid density profile $\rho/\rho_0$ showing the 
		vortex dipole in holography (left panels) and  
		DGPE (right panels) at four different times.
                Note the different subregions of the  ($x,y$)-grid.
We find good agreement
until shortly before the annihilation.
(a) Initial configuration: The vortices have a circular 
shape and are well separated. 
(b) Intermediate stage: The vortices approach each other and start to 
be deformed to an elliptical shape. 
(c) Shortly before annihilation: The vortices are strongly deformed and their density suppressions overlap. 
At this stage the dynamics can no longer be matched and the trajectories 
shown in \Fig{VortexTrajectories} end.
(d) After annihilation: shock waves propagate through the fluid and quickly decay, with 
clear differences in size and shape between holography and DGPE.
\label{fig:SnapshotsVortexDipole}
	}
\end{figure}

\begin{figure}[t]
	\includegraphics[width=0.9\columnwidth]{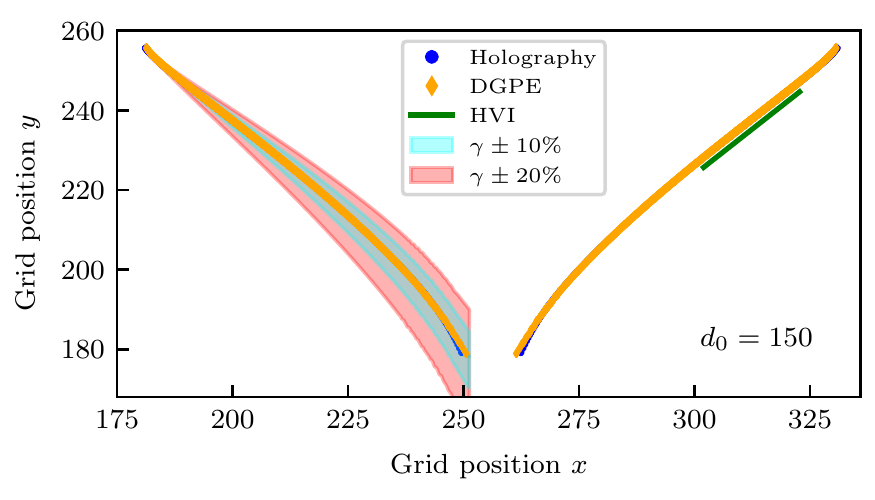}
	\caption{Vortex trajectories in holography for $\tilde{\mu}=6$ (blue dots) and in DGPE (orange diamonds) for an initial vortex dipole separation of $d_0 = 150$ grid points.
		Except just before the annihilation of the dipole, we find remarkable agreement between the trajectories. 
The error bands correspond to variations of the damping parameter $\gamma$ by $10\%$ (turquoise) and $20\%$ (red). 
We compare our data to the motion calculated from the HVI equations in a suitable (linear) regime (green solid line, shifted downwards by $-2.5$ grid points).
\label{fig:VortexTrajectories}
	}
\end{figure}

For the example of $\tilde{\mu} = 6$ and an initial 
separation of $d_0 = 150$ grid points, \Fig{SnapshotsVortexDipole} displays four characteristic 
snapshots of the superfluid density distributions in the holographic and DGPE simulations of the dipole. \Fig{VortexTrajectories} 
demonstrates that the matched vortex trajectories agree very well until shortly before the annihilation.
During the final stage of the annihilation process, the vortices deform and accelerate strongly.
We show the trajectories up to a few timesteps before the annihilation, at which point the matching procedure ceases to work.
The shock waves after the annihilation behave markedly different in the two theories. 

\Tab{DGPE_params} summarizes our results for the DGPE and HVI parameters matching the holographic dynamics for different $\tilde\mu$.
We note that the values of $\gamma$ are relatively large for the approximation underlying the DGPE. 
The DGPE can nevertheless be used as an effective phenomenological description of vortex dynamics, for which the consistent matching provides evidence. 
\begin{table}[t]
	\centering
	\caption{
          DGPE parameters (middle columns) for matching the holographic vortex shapes and trajectories 
          for various choices (left columns) of $\tilde\mu$, or equivalently $\tilde{T}/\tilde{T_\text{c}}$.
		The right columns show the friction coefficients, 
		$C = 1-(1+\gamma^{2})/[2\tau (1+\gamma^{2}|\ln\xi\,|^{2})]$ 
		and $C' =\gamma |\ln\xi\,| (1+\gamma^{2})/[2\tau (1+\gamma^{2}|\ln\xi\,|^{2})]$ of the
	   HVI equations, evaluated on a scaled grid with $\xi \to \xi/\hat{s}$. 
		We estimate the errors of all extracted parameters to be on the order of $1$-$2\%$.
	}
	\label{tab:DGPE_params}
	\begin{tabular}{C{0.7cm}C{1.2cm}|C{0.9cm}C{1.3cm}C{0.6cm}|C{0.9cm}C{0.9cm}C{0.9cm}C{0.01cm}}
		\hline \hline
		$\tilde{\mu}$& $\tilde{T}/\tilde{T_\text{c}}$& $\gamma$ & $\xi$   &$\tau$ & $\hat{s}$& $C$ & $C^{\prime}$ & \T\\
		\hline
		$4.5$ &0.9& $0.330$  & $9.25$  &  $2.85$ &  $114$   &$0.88$  & $0.095$  & \T  \\
		$6$ &0.68& $0.313$  & $3.50$&$ 5.76$ &  $116$  &$0.95$ & $0.047$ &\T \\
		$7.5$ &0.54& $0.297$  & $2.56$  & $7.46$ & $116$   & $0.97$ & $0.036$ &\T \\
		$9$ & 0.45&$0.281$  & $1.93$  &  $9.25$  & $119$ & $0.98$ & $0.029$ &\T\B\\
		\hline
	\end{tabular}
\end{table}

%
\textit{Dissipation} in the presence of vortices has a geometric interpretation in holography. Vortices correspond to tubes with vanishing $\Phi$ along their axis which punch holes through the charge cloud (see \Fig{BulkGeometry}) and allow for excitations in the boundary to fall into the black hole, thus dissipating energy to the black hole \cite{Chesler2013a.Science341.368}.
This favors the dissipation of ultraviolet modes, while the DGPE damps all modes except the zero-mode. 
The holographic dissipation through vortex cores naturally stops with the annihilation of the vortices. The observed difference in the arc waves after the vortex annihilation demonstrates a difference in the relative strength of dissipation with and without vortices in holography and DGPE. This suggests that the inherently strongly-coupled dissipation mechanism of the holographic framework is not fully captured by the DGPE. 

\textit{Real-world superfluids.} Having quantified the dissipation and friction parameters of the two-dimensional holographic superfluid, we compare them to experimentally accessible superfluids.

Ultracold, dilute Bose condensates, well described within the Gross-Pitaevskii framework, are typically prepared with alkali atoms. 
In experiments with a quasi two-dimensional trapping geometry,
disc-shaped clouds have been realized, with a thickness of $\sigma\sim10^{2}\dots10^{3}\,$nm \cite{Hadzibabic2009a,Johnstone2019a.Science.364.1267,Gauthier2019a.Science.364.1264}.
In such systems, 
collisions are captured
by the coupling $g = \sqrt{8 \pi} a / (M\sigma)$ \cite{Hadzibabic2009a,Petrov2000a.PhysRevLett.84.2551,Naidon2007a.NJP9.19}, 
where $a$ is the three-dimensional $s$-wave scattering length taking values of $a\simeq3\,$nm and $5\,$nm for $^{23}$Na and $^{87}$Rb, respectively.
Typical experimental surface densities of $\rho\sim10^{13}\,$m$^{-2}$ \cite{Hadzibabic2009a,Johnstone2019a.Science.364.1267,Gauthier2019a.Science.364.1264} 
lead to healing lengths on the order of $\xi=(4\sqrt{2\pi} a\rho/\sigma)^{-1/2}\sim0.4\dots 2\,\mu$m.
The quasi two-dimensional experimental settings satisfy the conditions for the applicability of the DGPE \eq{DGPE}
as the dimensionless parameter $\zeta_{\mathrm{2D}}\sim a/(\sqrt{2\pi}\sigma)$ is small. 
If the temperature $T$ of the condensed Bose gas is large compared to the zero-point energy but well below the critical temperature of the Berezinskii-Kosterlitz-Thouless transition \cite{Berezinskii1972JETP...34..610B,kosterlitz1973}, the damping $\gamma$ originates from the interactions between the condensed and thermal atoms.
This gives rise to a self-energy shift which in leading approximation implies $\gamma \simeq 12 M a^2 T/\pi$ \cite{Duine2004.PhysRevA.69.053623}.
For typical experimental temperatures of order $T \sim100\,$nK, one infers $\gamma\sim 10^{-4}\dots 10^{-3}$ 
which is two to three orders of magnitude smaller than the dissipation scale obtained 
above for the holographic model, see \Tab{DGPE_params}.
To induce such a strong dissipation
in an alkali gas would require increasing the scattering length, \eg, close to a Feshbach resonance while decreasing the density $\rho$ to
tune $\mu=\sqrt{8 \pi} a\rho / (M\sigma)$ and thus $\xi$ to the value matching the holographic model.
In the experimental setups realized so far, this requires $a\sim\sigma$.
Typically, such a system is difficult to be stabilized with bosons due to three-body-recombination loss prevailing at large scattering lengths \cite{Makotyn2014a.NatPhys.2.116}, which, however, can be remedied by using ultracold paired fermions \cite{Petrov2004.PhysRevLett.93.090404}.

It is interesting to note that friction coefficients $C'$ in the range quoted in our \Tab{DGPE_params} have been reported in thermally excited atomic Bose condensates at $T\sim300\,$nK, $C'\sim0.01\dots 0.03$ \cite{Moon2015a.PhysRevA.92.051601}. 

Finally, it is compelling to 
compare our results with measurements of vortex diffusivities in thin films of superfluid $^4\text{He}$ \cite{Kim1984a.PhysRevLett.52.53,Adams1987a.PhysRevB.35.4633,Harris2016a.NatPhys8.788,Sachkou2019a.Science366.1480}. 
Here, values of the HVI friction coefficient $C'$ very similar to our results for the holographic superfluid 
have indeed been measured in helium films at temperatures of order $T\sim1\,$K, $C'\sim10^{-2} \dots 1$ \cite{Kim1984a.PhysRevLett.52.53,Adams1987a.PhysRevB.35.4633,Finoretto1990a.PhysRevB.41.10994,Oda2009a.JLTP.158.262}. 
Note that in these experiments the temperature at which a certain $C'$ is observed also depends on the thickness of the film. 

\textit{Summary.}
Holography provides a higher-dimensional, field-theory based description of strongly dissipative superfluids. 
In this letter, we have performed a high-precision study of the dynamics of a vortex dipole. We have matched the quantum vortex dynamics of the DGPE as well as the vortex mechanics of the HVI equations to the two-dimensional holographic superfluid, thereby quantifying its dissipation.
The strongly dissipative character of the holographic superfluid is confirmed by the extracted values of the DGPE damping parameter $\gamma$ and of the HVI friction coefficient $C'$. 
Our findings suggest that holographic vortex dynamics can be applied to and be tested in experimentally accessible superfluids. 
Candidates for the experimental realization are strongly coupled Bose gases and, in particular, thin helium films with temperatures in the Kelvin range.
While we have derived these results from a simple vortex configuration, we expect holography to be applicable also to more complicated vortex ensembles, or turbulence in general, in strongly dissipative superfluids. 

\textit{Acknowledgments.} We thank C.~Barenghi, R.~Carretero-Gonz\'alez, M.~Karl, P.~G.~Kevrekidis and A.~Samberg for discussions and collaboration. This work was supported by EU Horizon-2020 (ERC Adv.~Grant EntangleGen, Project-ID 694561), by DFG (SFB 1225 ISOQUANT), and by Heidelberg University (CQD). 
P.\,W.\ was supported by the Studien\-stiftung des deutschen Volkes e.V. 


%

\clearpage
\begin{appendix}
\begin{center}
\onecolumngrid
\textbf{\large Appendix}
\end{center}
\setcounter{equation}{0}
\setcounter{table}{0}
\makeatletter

\section{Holographic superfluid: Equations of motion and their numerical solution} 
\label{app:EOMAdS/CFT}
%
In this appendix we explicitly give the equations of motion of the holographic model, discuss the boundary conditions for the scalar field $\Phi$ and for the gauge field $A_{\mu}$, and provide details of how to determine the boundary field configuration associated with the superfluid condensate.
See Ref.~\cite{Ewerz:2020} for a more detailed account.
In holography, a superfluid in two spatial dimensions has a dual gravity description in terms of an Abelian Higgs model on an asymptotically Anti-de Sitter spacetime in $3+1$ dimensions. 
It is described by the action (where not stated otherwise we use natural units where $\hbar, c, k_\mathrm{B}=1$)
\begin{align} \label{eq:ActionAdS/CFT}
S &= \frac 1{2 \kappa} \int d^4 x \sqrt{-\text{det}\,g_{\mu\nu}}\left (\mathcal{R} 
	- 2 \Lambda + \frac 1 {q^2} \mathcal{L}_{\text{gauge-matter}} \right), \nonumber \\
\mathcal{L}_{\text{gauge-matter}} 
   &= - \frac 1 4 F_{\mu \nu} F^{\mu \nu}- \lvert D_{\mu} \Phi \rvert^2 - m^2 \lvert \Phi \rvert^2\,.
\end{align}
Here, $\mu, \nu = t, x, y, z$ are the indices for the $(3+1)$-dimensional vector field. $ t, x, y$ denote the boundary coordinates, where we also use $\mathbf{r}=(x,y)$, and $z$ is the additional holographic coordinate.
$\Lambda = -3/L_{\mathrm{AdS}}^2$ is a negative cosmological constant, $L_{\mathrm{AdS}}$ is the curvature radius of the Anti-de Sitter spacetime, and $\kappa$ is Newton's constant in $3+1$ dimensions. 
$\mathcal{R}$ is the Ricci scalar of the metric $g_{\mu \nu}$.
The matter part is described by the field strength tensor $F_{\mu \nu} = \nabla_{\mu} A_{\nu} -  \nabla_{\nu} A_{\mu}$ with the gauge field $A_{\mu}$, and the associated gauge-covariant derivative $D_{\mu}= \nabla_{\mu} - i A_{\mu}$.
Note that in the holographic setting a $\lvert \Phi \rvert^4$-term is not required for symmetry breaking to occur \cite{Gubser:2008px}. 

We derive the holographic equations of motion in the probe limit  \cite{Hartnoll:2008vx}, \ie{}, we assume a large charge $q$ of the scalar field $\Phi$. 
Thus we can neglect the backreaction of the matter fields on the metric tensor $g_{\mu\nu}$. 
This is expected to be a good approximation at sufficiently high temperatures \cite{Albash:2009iq,Sonner2010a.PhysRevD.82.026001}.
Solving only the gravity part of the action \eq{ActionAdS/CFT} yields the $(3+1)$-dimensional Anti-de Sitter (AdS) spacetime metric which, using infalling Eddington-Finkelstein coordinates with respect to the holographic bulk direction $z$, takes the form
\begin{equation}\label{eq:metric}
	g_{\mu\nu}\text{d}r^{\mu}\text{d}r^{\nu}
	=\text{d}s^2
	=\frac{L_\text{AdS}^2}{z^2}\left(-h(z)\text{d}t^2+\text{d}x^2 +\text{d}y^2 -2\text{d}t\,\text{d}z\right)\,,
\end{equation}
where $h(z)=1-(z/z_\mathrm{h})^{3}$ is the horizon function of the planar Schwarzschild black hole at $z=z_\text{h}$.
Keeping this background metric fixed, the equations of motion for the gauge-matter part of the holographic model read
\begin{align}
\label{eq:EOMAdS/CFTApp}
\nabla_{\mu}F^{\mu \nu} 
= \i \left[\Phi^* D^{\nu} \Phi - \Phi \left ( D^{\nu} \Phi\right)^*\right]\,,
\qquad 
\left( D^2 + m^2 \right) \Phi 
= 0\,. 
\end{align}
We first solve the equations of motion for the background density, which is static and spatially homogeneous in the $(x,y)$-plane.
We therefore take the fields $\Phi$ and $A_{\mu}$ to be independent of the coordinates $x$, $y$ and $t$.
The gauge degree of freedom is fixed by choosing the axial gauge $A_z\equiv 0$.
Using the metric \eq{metric}, one finds the gauge field $A_\mu$ to obey the equations
\begin{align}
\label{eq:GaugeFields}
0 &= z^2 \partial_z^2 A_t + 2 \mathrm{Im}\left[\left(\partial_z \Phi\right)\Phi^*\right]\,,    \\
0 &= z^2 \left [h \partial_z^2 A_x +\left (\partial_z h \right) \left(\partial_z A_x \right)\right] - 2 \lvert \Phi \rvert^2 A_x\,,   \\
0 &= z^2 \left [h \partial_z^2 A_y +\left (\partial_z h \right) \left(\partial_z A_y \right)\right] - 2 \lvert \Phi \rvert^2 A_y\,,   \\
0 &=  2 A_t \lvert \Phi \rvert^2 - \i h \left[ \Phi^* \partial_z  \Phi - \left(\partial_z \Phi^*\right) \Phi \right]\,.
\end{align}
The last equation originates from the dynamic equation for $A_{z}$ and ensures the axial gauge fixing. 
The only dependence left in these equations is on the holographic coordinate $z$.
The equation for the scalar field reads
\begin{equation}
\label{eq:ScalarField}
0 
=  z^2 h \partial_z^2 \Phi 
    - z \left ( -2 \i z A_t  + 2h -z \partial_z h \right)\partial_z \Phi 
    - \left(2 \i z A_t - \i z^2 \partial_z A_t + z^2 \mathbf{A}^2 +m^2 \right) \Phi\,, 
\end{equation}
where $\mathbf{A} = (A_x, A_y)$.
The thermal-equilibrium configuration of the condensate field $\psi$ can then be obtained by solving Eqs.~\eq{GaugeFields}--\eq{ScalarField} and expanding the solution for the scalar field near the boundary, $z\to 0$, according to the holographic dictionary, 
\begin{align} \label{eq:BoundaryExpansion}
	\Phi(t, \bm{r}, z)=\eta(t, \bm{r})\,z +  \psi(\bm{r},t) \,z^2 +\mathcal{O}(z^3)\,,
\end{align}
where $\eta(t, \bm{r})$ is set to zero by choosing corresponding boundary conditions.

We imprinted vortices onto this background field configuration as described in detail in \App{MatchingProfiles}.
Their subsequent time evolution was computed by solving the full set of dynamical equations \eq{EOMAdS/CFTApp} without the  assumptions of homogeneity in $\mathbf{r}$ and stationarity. 
For numerical purposes it turns out to be convenient to rescale the scalar field as $\tilde{\Phi} = \Phi/z$ and rewrite the equations in terms of the `lightcone derivative' 
\begin{equation}
\nabla_+ X = \partial_t X - \frac{h(z)}{2} \,\partial_z X
\end{equation} 
of the fields $X \in\{ A_x, A_y, \tilde{\Phi}\}$. 
The resulting equations of motion read
\begin{align}\label{eq:FullEoMHolo}
\partial_z^2 A_t &= \partial_z \mathbf{\nabla} \cdot \mathbf{A} - 2 \mathrm{Im}\left(\tilde{\Phi}^*\partial_z \tilde{\Phi}\right)\,, \\
\partial_z \nabla_+ A_x &= \frac 1 2 \left[\partial_y^2 A_x + \partial_x \left(\partial_z A_t - \partial_y A_y \right) \right] - \lvert \tilde{\Phi}\rvert^2 A_x
+ \mathrm{Im}\left(\tilde{\Phi}^*\partial_x \tilde{\Phi}\right)\,, \\
\partial_z \nabla_+ A_y &= \frac 1 2 \left[\partial_x^2 A_y + \partial_y \left(\partial_z A_t - \partial_x A_x \right) \right] - \lvert \tilde{\Phi}\rvert^2 A_y
+ \mathrm{Im}\left(\tilde{\Phi}^*\partial_y \tilde{\Phi}\right)\,, \\
\partial_z \nabla_+ \tilde{\Phi} &= \frac 1 2 \mathbf{\nabla}^2 \tilde{\Phi} - \i \mathbf{A} \cdot \mathbf{\nabla}\tilde{\Phi} + \i A_t \partial_z \tilde{\Phi} - \frac \i 2 \left(\mathbf{\nabla} \cdot \mathbf{A} - \partial_z A_t \right) \tilde{\Phi} - \frac 1 2 \left(z  + \mathbf{A}^2 \right) \tilde{\Phi}\,,
\end{align}
where $\mathbf{\nabla} = (\partial_x, \partial_y)$.
The static background solution representing the thermal-equilibrium condensate as well as the solutions to the dynamical equations of motion are subject to the following boundary conditions (see, \eg{}, Ref.~\cite{Chesler2013a.Science341.368}):
\begin{align}
A_t (z= 0) &= \tilde{\mu}\,, \, &A_t(z=\zh) = 0\,, \\
A_x (z= 0) &= 0\,, \, &A_x(z=\zh) = 0\,, \\
A_y (z= 0) &= 0\,, \, &A_y(z=\zh) = 0\,, \\
\partial_z \Phi (z) \lvert_{z=0} &= 0\,.
\end{align}
In addition, the scalar field $\Phi$ needs to be regular at the horizon $z=\zh$ \cite{Son2002a.JHEP.2002.042}, corresponding to infalling boundary conditions.  
In the above equations, all holographic fields are measured in units specified by setting $\zh =1$.
We furthermore choose the curvature radius to be $L_{\mathrm{AdS}}=1$ and take the mass of the scalar field to be $m^2 = -2/L_{\mathrm{AdS}}^2 = -2$ which is well above the Breitenlohner-Freedman bound $-9/4$ in $3+1$ dimensions thus ensuring stability of $\Phi$ against tachyonic decay \cite{Breitenlohner1982a.PLB.115.197,Breitenlohner1982b.AP.144.249}.

In the dual interpretation, the planar black hole at $z=z_\mathrm{h}$, in the probe limit, corresponds to a static heat bath with temperature $\tilde{T} = 3/(4 \pi z_\text{h})= 3/(4 \pi)$ in our units \cite{Hartnoll:2008vx,Herzog:2008he}.
In the probe limit, the only free parameter left is the chemical potential $\tilde\mu$ or, equivalently, the dimensionless reduced temperature $\tilde{T}/\tilde{\mu}$.
The critical chemical potential at which the order parameter $\psi$ vanishes, in our units $\tilde\mu_\mathrm{c}\simeq4.064$, fixes the combination $\tilde{\mu}\tilde{T}/\tilde{T}_{\mathrm{c}}=\tilde\mu_\mathrm{c}$, where $\tilde{T}_\text{c}$ denotes the critical temperature \cite{Hartnoll:2008vx,Herzog:2008he,Sonner2010a.PhysRevD.82.026001,Anninos:2010sq}.
Hence, within the range of values we chose for the chemical potential, $\tilde{\mu}=4.5\dots9$, the system is in the superfluid phase, \ie{}, $\tilde{T}/\tilde{T}_{\mathrm{c}}=\tilde\mu_\mathrm{c}/\tilde\mu\simeq0.45\dots0.9$, \cf{} \Tab{DGPE_params}  for four different examples.

All numerical computations were performed on a $512\times512$  grid in the $(x, y)$-plane, imposing periodic boundary conditions. 
The grid spacing is always chosen as $\ell=1/8$.
In the holographic $z$-direction, we used 32 Chebyshev polynomials, while for the time evolution a fourth-fifth order Runge-Kutta-Fehlberg algorithm with adaptive time steps was used.
In this scheme one unit of time is composed of 10 to 1000 numerical timesteps. 
More details on our numerical evaluation of the holographic equations are given in Ref.~\cite{Ewerz:2020}.

\section{Dissipative Gross-Pitaevskii equation and its implications for vortex motion} 
\label{app:DGPE}

\subsection{The dissipative Gross-Pitaevskii model and its application to ultracold dilute Bose systems}
\label{app:DGPEParams}
%
In this section, we summarize the implications of the Gross-Pitaevskii (GP) model for the description of  cold Bose systems. 
For a concise discussion see, \eg, Ref.~\cite{Proukakis_2008}.
Reinstating $\hbar$, the DGPE \eq{DGPE} reads
\begin{equation} \label{eq:app:DGPE}
\hbar\,\partial_t \psi(\mathbf{r},t) = - \left( {\i + \gamma} \right) \left [ - \frac {\hbar^{2}}{2M }  {\nabla^2} + g\, \lvert \psi (\mathbf{r},t) \vert^2 - \mu \right] \psi(\mathbf{r},t) \, .
\end{equation}
$M$ is the mass of the bosons, $g$ characterizes their interactions, $\gamma$ is a dimensionless phenomenological damping parameter quantifying the dissipation, and $\mu$ is a chemical potential representing a constant shift of the single-particle energy. The DGPE \eq{app:DGPE} has the same form for a Bose condensate in two and three spatial dimensions, and we will in the following consider both cases. In this appendix, we will therefore treat $\mathbf{r}$ as a two- or three-dimensional position vector depending on the context. 

For matching the GP vortex dynamics to that in the holographic superfluid, we numerically solved 
\Eq{app:DGPE} in two spatial dimensions, as we will discuss in more detail in \App{dlDGPE} below. 
One goal of the matching procedure was to extract the parameter $\gamma$, and subsequently, in a point-particle picture for vortices, phenomenological coefficients quantifying the mutual friction between the vortices and the superfluid.
To set the stage for this matching we briefly comment, in the following, on the conditions under which \Eq{app:DGPE} is applicable.
Moreover, for a comparison of the numerics with results obtained in experiments performed with \emph{quasi two-dimensional} (2D) trapping potentials, we need to distinguish different conditions in the crossover regime between strictly two and three spatial dimensions.

Quite generally, the GP model provides a quantitative description of dilute ultracold, \ie{} condensed -- or `degenerate' -- Bose gases, typically prepared with alkali atoms.
Dilute means that the length scale characterising the collisional interactions is much smaller than the mean interparticle spacing.
While at very low temperatures ($k_{B}T\ll\mu$) the GP equation itself provides a good approximation of the condensate dynamics, a self-consistent evaluation of the combined dynamics of the condensate field $\psi$ and the thermal component is required at higher temperatures and is increasingly difficult the closer $T$ is to the critical temperature of Bose-Einstein condensation.
An even more intricate problem is set by non-dilute systems such as superfluid helium where strong quantum fluctuations and depletion of the condensate also at very low temperatures limit the applicability of \Eq{app:DGPE} and one generically resorts to other approaches, in particular the Tisza-Landau two-fluid model \cite{Tisza1938TPiHII,PhysRev.60.356,Donnelly2005a}.

The atoms' interactions at the energies prevailing in dilute systems are well captured by a single parameter, the $s$-wave scattering length $a$, with values of $\simeq 3\,$nm and $\simeq 5\,$nm for the most commonly used elements ${}^{23}$Na and ${}^{87}$Rb, respectively.
The $s$-wave approximation implies that the two-body interactions entering the many-body field theory are local in space and time, meaning that the ultraviolet length scale below which this locality is violated corresponds to collision energies much higher than those attained in the low-temperature system, see, \eg{}, \cite{Bloch2008a.RevModPhys.80.885}.

However, the actual value of the respective GP coupling parameter $g$, and the conditions for mean-field and perturbative approximations to be valid, depend on the dimensionality of the system.
Let us, for simplicity, assume that the Bose system described by \Eq{app:DGPE} is confined within a $d$-dimensional box of volume $\mathcal{V}$ with periodic boundary conditions such that its ground state is characterized by a uniform mean-field condensate density $\rho_0=\mathcal{V}^{-1}\int_{\mathcal{V}} d\mathbf{r}\,|\psi(\mathbf{r},t)|^{2}\equiv\|\psi\|^{2}$.
For gases trapped in three-dimensional volumes $\mathcal{V}$ ranging between $10^{-7}$ and $10^{-10}\,$cm$^{3}$, typical experimental densities $\rho_\mathrm{3D}$ are between $10^{12}$ and $10^{15}$ particles per cm$^{3}$ \cite{Bloch2008a.RevModPhys.80.885}. 
In three spatial dimensions, the interactions are characterized by the coupling constant $g=g_{\mathrm{3D}}=4\pi\hbar^{2}a/M$, with the three-dimensional (3D) $s$-wave scattering length $a$.
For a given coupling $g$ and a condensate density $\rho$ in $d$ dimensions, the characteristic length scale of the condensate is the healing length $\xi = \hbar/(2M g\rho_0)^{1/2}$. 
$\xi$ is the scale on which the density rises to the uniform background value $\rho_{0}$ near an infinite potential wall or which determines the diameter of the core of a vortex. 

An important parameter is the 3D diluteness $\zeta=\sqrt{\rho_\mathrm{3D} a^{3}}$, which depends on the ratio between $a$ and the mean interparticle separation and which needs to be small for a mean-field description in terms of \Eq{app:DGPE} alone or including perturbative corrections to \eq{app:DGPE}  to be valid in three spatial dimensions.
In dilute alkali gases it is $\zeta\sim10^{-3}$ while in superfluids like $^{4}$He it is of order unity. 

More relevant for the settings discussed in this letter, various experimental realizations with a quasi-2D trapping geometry have been achieved, with a transverse confinement down to a cloud thickness of $\sigma\sim10^{2} \dots 10^{3}\,$nm \cite{Hadzibabic2008a.NJP.10.045006,Hadzibabic2009a,Neely2012a.PhysRevLett.111.235301,Kwon2014a.PhysRevA.90.063627,Moon2015a.PhysRevA.92.051601,Johnstone2019a.Science.364.1267,Gauthier2019a.Science.364.1264}.
As this thickness is still considerably larger than the respective 3D scattering length $a$, the coupling entering the DGPE \eq{app:DGPE} in a quasi-2D setting is given by $g = \sqrt{8 \pi}\hbar^{2} a / (M\sigma)$ \cite{Hadzibabic2009a,Petrov2000a.PhysRevLett.84.2551,Naidon2007a.NJP9.19}.
While generic settings are characterized by the small dimensionless ratio $a/\sigma$, substantially larger values of $g$ are possible near a so-called confinement-induced resonance \cite{Petrov2001a.PhysRevA.64.012706}.
Typical experimental 2D densities are $\rho\sim10^{13}\,$m$^{-2}$ \cite{Hadzibabic2009a,Johnstone2019a.Science.364.1267,Gauthier2019a.Science.364.1264}, implying healing lengths $\xi=(4\sqrt{2\pi} a\rho/\sigma)^{-1/2}$ between $0.4$ and $2\,\mu$m, which are usually smaller than or similar to the cloud thickness $\sigma$.
Such cloud geometries are still 3D for the collisions between particles. 
Hence, if also the particle motion in the confined direction occupies many trap levels, the validity of mean-field and perturbative approximations is ensured by a small 3D diluteness parameter $\zeta = \sqrt{\rho a^3 / \sigma}$, with 2D density $\rho$.

In typical quasi-2D experimental settings, however, only the lowest transverse trap mode is populated such that perturbation theory is controlled by the small dimensionless quantity $\zeta_{\mathrm{2D}}= a/(\sqrt{2\pi}\sigma)$ which takes values between $10^{-2}$ and $10^{-3}$ and does not depend on the particle density \cite{Hadzibabic2009a,Johnstone2019a.Science.364.1267,Gauthier2019a.Science.364.1264,Bloch2008a.RevModPhys.80.885}.
For tighter confinement, \ie, approaching the 2D limit, $\zeta_{\mathrm{2D}}$ becomes sensitive to the density, such that the expansion parameter is given by $\zeta_{\mathrm{2D}}\sim1/|\ln(\rho a_{\mathrm{2D}}^{2})|$. 
Here, $a_{\mathrm{2D}}=\sigma\sqrt{\pi/B}\exp\{-\sqrt{\pi/2}\,\sigma/a\}$, with Catalan constant $B \simeq 0.916$, is the respective 2D scattering length, \ie{}, $\zeta_{\mathrm{2D}}\sim1/|\ln(\rho \sigma^{2}\pi/B)-\sqrt{2\pi}\,\sigma/a|$, which exhibits the confinement-induced resonance  \cite{Petrov2001a.PhysRevA.64.012706,Hadzibabic2009a,Naidon2007a.NJP9.19,Mora2009.PhysRevLett.102.180404}.

We finally briefly comment on the dissipative damping of the condensate field $\psi$ caused by the interactions, which is taken into account in the DGPE \eq{app:DGPE} by a non-zero value of the dimensionless phenomenological damping parameter $\gamma$.
For further discussion in the context of vortex dynamics see Apps.~\app{HVI} and \app{Diffusivity}.

If the temperature $T$ of the condensed Bose gas is large compared to the zero-point energy but well below the critical temperature of the phase transition, which in 2D is of Berezinskii-Kosterlitz-Thouless (BKT) type, the interactions between the condensed and thermal particles give rise to exponential damping in the limit of long evolution times, once initial effects have been damped out which may arise, \eg{}, from the particular quench bringing the system out of equilibrium. 
Within a perturbative expansion of the time-dependent self-energy, the damping parameter $\gamma$ results, in two-particle-irreducible two-loop approximation, quadratic in $g$ and thus $a$, as \cite{Duine2004.PhysRevA.69.053623}
\begin{align}
\gamma\simeq12\,Ma^{2}k_{B}T/(\pi\hbar^{2})\,.
\label{eq:gammaDuine}
\end{align}
We note that, if we  take the chemical potential $\mu$ in \Eq{app:DGPE} to equal the energy eigenvalue $\mu = g\rho_0$ of the zero-momentum eigenstate of the stationary GPE for a homogeneous gas at zero temperature, all but the zero-mode are damped for $\gamma>0$.
Hence, obtaining self-consistent stationary solutions for a condensate at non-vanishing temperatures in general requires to go beyond the DGPE \eq{app:DGPE}. 
One possibility is the inclusion of a noise term and thus the extension of \eq{app:DGPE} to a stochastic differential equation \cite{Duine2001a.PhysRevA.65.013603,Cockburn2013qgft.conf}, or the coupling of the GPE to the dynamics of higher-order correlators \cite{Zaremba1999a.JLTP116.277,Allen2013a} which account for a self-consistent treatment of condensed and non-condensed particles.
If the temperature is too high for the perturbative estimate \eq{gammaDuine} of $\gamma$ to apply, such methods have been used to obtain an estimate, see the discussion in \App{Diffusivity} below.

But even in these cases, the DGPE can provide a quantitatively good description of dissipative dynamics as long as the system is away from stationarity and as long as the thermal fraction, which gives rise to the dissipation, does not become significantly disturbed by the evolving condensate field. 
As we are, here, primarily interested in the matching of the vortex dynamics subject to friction exerted by a static thermal bath with that obtained in the holographic framework, we make use of \Eq{app:DGPE} as a phenomenological description, which we use to determine $\gamma$ through the matching and compare it with experimentally obtained values as well as other theoretical approaches.

\subsection{Units and numerical solution}
\label{app:dlDGPE}
%
We numerically solve a dimensionless form of the DGPE \eq{app:DGPE} which is obtained by rescaling the physical parameters with respect to some length scale $\lengthscale$ which sets, together with the mass $M$ and $\hbar$, both the spatial and temporal units.
For this, one introduces dimensionless primed quantities by defining
$\mathbf{r}=\dimless{\mathbf{r}}\lengthscale$, $t=\dimless tM\lengthscale^{2}/\hbar$,  $g=\dimless g\hbar^{2}\lengthscale^{d-2}/M$, $\mu=\dimless\mu\hbar^{2}/(M\lengthscale^{2})$.
Here, we work in two spatial dimensions such that $d=2$.
We furthermore define the dimensionless and normalized complex field $u({\mathbf{r}}, t)$ by $\psi(\mathbf{r},t)=\sqrt{\rho_{0}}\,u({\mathbf{r}}, t)$, with the background density $\rho_0=\dimless{\rho}_0\lengthscale^{-2}$.

Alternatively, we can make use of the freedom to choose the temporal and spatial grid units independently.
For this, we introduce the dimensionless coordinate and time (denoted with overbar) as ${\mathbf{r}}=\gridunit{\mathbf{r}}s$ and $t=\gridunit{t} M s^{2}/\hbar\tau$ in terms of the spatial and temporal lattice units $s$ and $s^{2}/\tau$, respectively, which corresponds to a rescaling $\lengthscale=\gridlengthscale s$ and thus $\gridunit{\mathbf{r}}=\dimless{\mathbf{r}}\gridlengthscale$, etc., while $\gridunit{t}=\dimless t \,\gridlengthscale^{2}\tau$.
Setting, furthermore, $\bar\mu=\bar g\bar\rho_0=\dimless\mu/\gridlengthscale^{2}$ and suppressing the overbar on all quantities measured in grid units, yields the dimensionless DGPE in the form
\begin{equation} \label{eq:dlDGPE}
\partial_t u(\mathbf{r},t) = \frac {\i + \gamma} {2\tau} \left [  {\nabla^2} + 2\mu\left(1- \lvert u(\mathbf{r},t) \vert^2 \right) \right] u(\mathbf{r},t)
\end{equation}
for the complex field $u$. 
The dimensionless healing length is given by $\xi=(2\mu)^{-1/2}$.

The DGPE parameters $\gamma$, $\tau$, and $\xi$ are then adjusted to match the characteristics of the vortices and their dynamics observed in the holographic calculations, see \Tab{DGPE_params}.
To ensure that both the holographic as well as the DGPE systems are simulated with the same numerical resolution, we  use the same spatial grids with $512 \times 512$ points in the $(x, y)$-plane and an equal number of evaluation points along the time direction.
In numerically solving \Eq{dlDGPE} we make use of a spectral split-step algorithm with fixed time step on high-level graphical processing units.

\subsection{Vortex motion in the dissipative system} 
\label{app:HVI}
%
The motion of a vortex dipole as we consider it here has been explored separately in holographic and DGPE simulations in \cite{Lan:2018llf,Ewerz:2020} and \cite{Lan2020.arXiv.2003.01376}, respectively.
More generally, the motion of vortices in a two-dimensional system and their interactions with each other and with background excitations has been studied in depth, both theoretically and experimentally, also in the vicinity of the BKT transition, in thin films of superfluid helium \cite{Hall1956a.PRSLA.238.204,Iordansky1964a.AnnPhys.29.335,Iordanskii1966JETP...22..160I,Ambegaokar1978a.PhysRevLett.40.783,Sonin1987a.RevModPhys.59.87,Ambegaokar1980a,Sonin1997a.PhysRevB.55.485,Thouless1996a.PhysRevLett.76.3758,Thompson2012a.PhysRevLett.108.184501,Thompson2013a.JLTP.171.459,Donnelly2005a,Kim1984a.PhysRevLett.52.53,Adams1987a.PhysRevB.35.4633,Harris2016a.NatPhys8.788,Sachkou2019a.Science366.1480}, in superconductors, \eg{}, \cite{Blatter1994a.RevModPhys.66.1125}, as well as in ultracold atomic gases \cite{Rosenbusch2002a.PhysRevLett.89.200403,AboShaeer2002.PhysRevLett.88.070409,Hadzibabic2008a.NJP.10.045006,Hadzibabic2009a,Henn2009a.PhysRevLett.103.045301,Neely2012a.PhysRevLett.111.235301,Kwon2014a.PhysRevA.90.063627,Moon2015a.PhysRevA.92.051601,Johnstone2019a.Science.364.1267,Gauthier2019a.Science.364.1264}.

Much focus has been set on the motion of vortices as point particles as described by the Hall-Vinen-Iordanskii (HVI) equations \cite{Hall1956a.PRSLA.238.204,Iordansky1964a.AnnPhys.29.335,Iordanskii1966JETP...22..160I}.
In their simplest form, they describe the mechanical motion 
of $N$ vortices due to the different forces exerted by the fluid.
The HVI equations determine the velocity of the $i$-th vortex at position $\mathbf{r}_{i}$  $(i=1,\dots,N)$,
\begin{align}
\mathbf{v}_{i}=
\frac{\mathrm{d}\mathbf{r}_i}{\mathrm{d}t} 
&= \mathbf{v}_\mathrm{s}^{i} + C(\mathbf{v}_\mathrm{n} - \mathbf{v}_\mathrm{s}^{i} )  
    + w_i  \, C' \hat{\mathbf{e}}_\perp\times (\mathbf{v}_\mathrm{n} - \mathbf{v}_\mathrm{s}^{i})\,,
\label{eq:HVIfull}
\end{align}
where $\mathbf{v}_\mathrm{s}^{i}$ is the superfluid velocity created by all vortices $j\not=i$, $\mathbf{v}_\mathrm{n}$ is the velocity of the normal fluid, $w_{i}=\pm1$ is the winding number of the vortex, $\hat{\mathbf{e}}_\perp$ is a unit vector perpendicular to the $(x,y)$-plane in a right-handed coordinate system, and $C$ and $C'$ are phenomenological friction coefficients.
All quantities are taken to be dimensionless in the units of \Eq{dlDGPE}.
\Eq{HVIfull} is obtained by balancing the Magnus force resulting from the vortex moving relative to the superfluid  \cite{Hall1956a.PRSLA.238.204} against the drag forces along and perpendicular to the relative velocity of the vortex and the fluid, and by solving for the vortex velocity \cite{Ambegaokar1980a,Sonin1997a.PhysRevB.55.485}.

As we consider the dynamics of the superfluid component of the system only, we neglect the normal-fluid velocity $\mathbf{v}_\mathrm{n}$.
As a result, the HVI equations take the simpler form
\begin{align}
\mathbf{v}_{i}=
\frac{\mathrm{d}\mathbf{r}_i}{\mathrm{d}t} 
&= (1-C)\,\mathbf{v}_\mathrm{s}^{i}   
    - w_i  \, C' \,\hat{\mathbf{e}}_\perp\times \mathbf{v}_\mathrm{s}^{i}\,.
\label{eq:HVIsf}
\end{align}
In this case the friction between the vortices and the superfluid reduces,  by a factor $1-C$, the velocity $\mathbf{v}_{i}$ of the vortex relative to the velocity $\mathbf{v}^{i}_\mathrm{s}$ of the superfluid created by the other vortices.
Moreover, this velocity difference gives rise to a Magnus force proportional in magnitude and perpendicular to $\mathbf{v}_{i}-\mathbf{v}^{i}_\mathrm{s}$. 

The motion of the $i$-th vortex is then solely determined by the superfluid velocity created by the other vortices in the system, 
\begin{align}
\mathbf{v}_\mathrm{s}^{i}
={2\pi w_{i}}\,\hat{\mathbf{e}}_\perp\times\nabla_{\mathbf{r}_{i}}H_\mathrm{PV}(\{\mathbf{r}_{k},w_{k}\})
=2\sum_{\{j\,|\,j\not=i\,\}}w_{j}\frac{\hat{\mathbf{e}}_\perp\times(\mathbf{r}_{i}-\mathbf{r}_{j})}{|\mathbf{r}_{i}-\mathbf{r}_{j}|^{2}} \,,
\label{eq:OnsagerVelocities}
\end{align}
which is derived from the Kirchhoff-Onsager point-vortex Hamiltonian $H_\text{PV}$ \cite{Onsager1949a},
\begin{align}
H_\mathrm{PV}(\{\mathbf{r}_{k},w_{k}\,|\,k=1,\dots, N\})=(2\pi)^{-1}\sum_{i\not=j}w_{i}w_{j}\ln|\mathbf{r}_{i}-\mathbf{r}_{j}|\,.
\label{eq:POEnergy}
\end{align}
The theoretical calculation and understanding of the drag entering \Eq{HVIfull} has been the subject of extensive work, see \eg{}~\cite{Hall1956a.PRSLA.238.204,Iordansky1964a.AnnPhys.29.335,Iordanskii1966JETP...22..160I,Ambegaokar1978a.PhysRevLett.40.783,Sonin1987a.RevModPhys.59.87,Ambegaokar1980a,Sonin1997a.PhysRevB.55.485,Thouless1996a.PhysRevLett.76.3758,Duine2004.PhysRevA.69.053623,Thompson2012a.PhysRevLett.108.184501,Thompson2013a.JLTP.171.459,Donnelly2005a} and references therein. 
Here, we are primarily interested in the relation between the HVI equations and the phenomenological description of the superfluid dynamics in terms of the DGPE.
The DGPE \eq{dlDGPE} is a different form of the Complex Ginzburg-Landau equation (CGLE) \cite{Aranson_2002} 
\begin{equation}
	\left(\delta_{\xi} + \i\alpha \right) \partial_t u_\xi 
	= \nabla^{2} u_\xi + \frac{1}{\xi^2}\left(1- |u_\xi|^2\right)u_{\xi}\,. 
	\label{eq:dlDGPEmiot}
\end{equation}
As was shown in the context of the CGLE in \cite{Miot2009a.AnalPDE.2.159,Kurzke2009a.AnalPDE.2.159}, in the point-vortex limit, \ie, for vortex core sizes much smaller than the distances $|\mathbf{r}_{i}-\mathbf{r}_{j}|$ between the defects, the above HVI equations provide an approximate description of the motion of vortices resulting from \eq{dlDGPEmiot}. Its solutions $u_{\xi}$ depend on the parameters $\xi$ and $\delta_{\xi}$.   
Solutions $u_{\xi}$ containing $N$ vortices at positions $\mathbf{r}_{i}$ and with winding numbers $w_{i}$ can be written as
\begin{equation}
	u_\xi (\mathbf{r}_i, w_i;N)(\zc) 
	= \prod_{i=1}^{N} f_{1,w_{i}}\left(\frac{|\zc - \zc_i|}{\xi}\right)\,\left(\frac{\zc-\zc_i}{|\zc - \zc_i|}\right)^{w_i}\,,
	\label{eq:VortexSol}
\end{equation}
where $\mathbf{r}_i=(x_i,y_i)$, $\zc=x + \i y$, $\zc_i = x_{i} + \i \, y_{i}$, and where the function $f_{1,w}:\mathbb{R}^{+}\to[0,1]$, with $f_{1,w}(0)=0$ and $f_{1,w}(\infty)=1$, describes the vortex core profile in the case $\xi=1$.
As before, $\xi$ controls the core size of the vortices in the solutions of \Eq{dlDGPEmiot}.
In the asymptotic limit of $\xi\to0$, with $\delta_{\xi}|\ln\xi\,|\equiv\delta\in(0,\infty)$ kept fixed, the motion of the vortices \eq{VortexSol} according to the CGLE \eq{dlDGPEmiot} is well described by the HVI equations \eq{HVIsf} with friction coefficients
$C$ and $C'$ defined by
\begin{align}
C = 1+\frac{\alpha}{\alpha^{2}+\delta^{2}}\,,
\qquad 
C'=\frac{\delta}{\alpha^{2}+\delta^{2}}\,.
\label{eq:DragCoeffCGL}
\end{align}
The parameters of the CGLE \eq{dlDGPEmiot} are related to those of the DGPE \eq{dlDGPE} by
\begin{align}
\delta_{\xi} = \frac{2\tau\gamma}{1+\gamma^{2}}\,,
\qquad 
\alpha=-\frac{2\tau}{1+\gamma^{2}}\,.
\label{eq:CGLvsDGPECoeff}
\end{align}
Note that in \cite{Miot2009a.AnalPDE.2.159}, $\alpha$ was set to $1$, which hence is equivalent to switching the sign of time and rescaling time in \Eq{dlDGPE} by choosing $\tau=-(\gamma^{2}+1)/2$, which gives $\delta_{\xi}=-\gamma$.
Hence, the above definitions generalize the expressions given in \cite{Miot2009a.AnalPDE.2.159} to the case $\alpha\not=1$, \ie{}, they allow us to choose the temporal and spatial grid units independently, as discussed in \App{dlDGPE}.

Considering the force balance which yields the HVI equations shows that, in units where $\alpha=1$, the parameter $\delta$ equals the ratio of the coefficient $C'$ of the Magnus force, which appears when the vortex velocity differs from the superfluid velocity \cite{Sonin1997a.PhysRevB.55.485}, and the negative of the respective coefficient $1-C$ of the friction force between vortex and superfluid, that is $\delta = C'/(C-1)$. 
In a holographic superfluid, the Magnus force on a vortex has been shown to serve as a probe for the charge density outside the horizon \cite{Iqbal2012a.CQG.29.194004}.

We finally obtain, with $\delta=\delta_{\xi}|\ln\xi\,|$, the asymptotically $(\xi\to0)$ valid coefficients
\begin{align}
C = 1-\frac{1}{2\tau}\frac{1+\gamma^{2}}{1+\gamma^{2}|\ln\xi\,|^{2}}\,,
\qquad 
C'=\frac{\gamma|\ln\xi\,|}{2\tau}\frac{1+\gamma^{2}}{1+\gamma^{2}|\ln\xi\,|^{2}}\,.
\label{eq:DragCoeffDGPE}
\end{align}

For the case of a vortex dipole, the HVI equations \eq{HVIsf} can be integrated easily to provide expressions for the vortex trajectories, which are found to be straight lines. 
Defining, for $N=2$ vortices with opposite winding numbers $w_{1}=-w_{2}$, the relative and center coordinates
\begin{align}
\mathbf{d}(t) = (\mathbf{r}_{1}-\mathbf{r}_{2})(t)=d(t)\,\hat{\mathbf{d}}_{0}\,,\qquad
\mathbf{R}(t) = (\mathbf{r}_{1}+\mathbf{r}_{2})(t)/2=R(t)\,\hat{\mathbf{d}}_{0}\times\hat{\mathbf{e}}_\perp\,,
\label{eq:DdPair}
\end{align}
with $\hat{\mathbf{d}}_{0}=\mathbf{d}(0)/|\mathbf{d}(0)|$, $d_{0}=d(0)$, one finds the solutions
\begin{equation}
	d(t) = \sqrt{d_{0}^2 - 8 \, C' \,t}\,,
	\qquad
	R(t) = R(0) + w_{1}\frac{1-C}{2C'}\left[d_{0} -d(t)\right]\,.
	\label{eq:doftRoft}
\end{equation}
Besides the dissipative motion considered here, the same scaling law for the approach of vortices, $d(t)\sim\sqrt{t_{I}-t}$ with $t_I = d_0^2/(8C')$, also applies to the evolution of vortex lines in three-dimensional superfluids as studied with vortex filament models \cite{Schwarz1988a.PhysRevB.38.2398,Waele1994a.PhysRevLett.72.482}, GPE simulations \cite{Koplik1993a.PhysRevLett.71.1375,Nazarenko2003a.JLTP.132.1,Tebbs2011a.JLTP.162.314}, and as observed in experiments \cite{Bewley2008a.PNAS.105.13707,Paoletti2010a.PhysicaD239.1367}.
Note that, in 3D, this law also applies to non-dissipative, purely Hamiltonian evolution where the vortex lines undergo an instability when approaching each other \cite{Koplik1993a.PhysRevLett.71.1375,Tebbs2011a.JLTP.162.314}, analytically shown in the linear regime where the cores start to overlap \cite{Nazarenko2003a.JLTP.132.1}. 

\begin{figure}
	\begin{center}
		\includegraphics[width=0.5\textwidth]{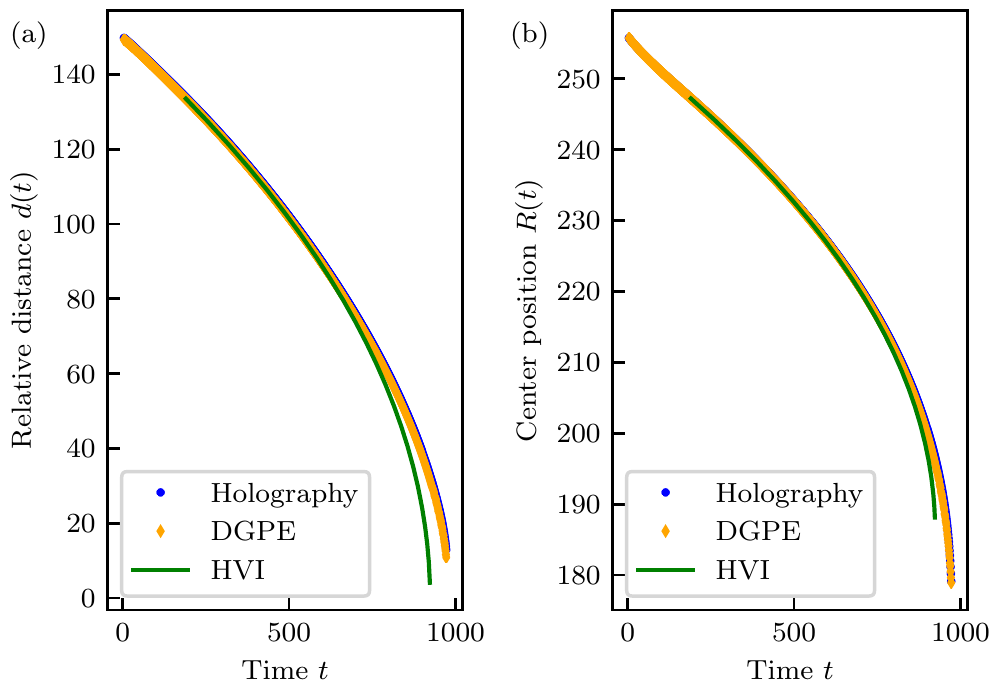}
		\caption{Temporal evolution of (a) the relative distance $d(t)$ and (b) the center position $R(t)$ of a vortex dipole with initial separation $d_0 = 150$ as obtained from the holographic model (blue dots), the DGPE (orange diamonds) and the HVI equations (green solid line, \cf{}~\Eq{doftRoft}). 
		All quantities are measured in grid units as defined in \App{dlDGPE}.
		The starting time of the HVI curves is the same here and in \Fig{VortexTrajectories}.
		Up to time $t \simeq 600$, the HVI solution agrees well with the results from 
		the DGPE and the holographic model.
		For $t \gtrsim 600$, the HVI equations predict a faster approach of the vortex-antivortex pair resulting in a shorter lifetime of the vortex dipole.
		Note that the last point of the HVI trajectory shown in \Fig{VortexTrajectories} 
		corresponds to $t = 575$. 
				\label{fig:RtAnddt}
		}
	\vspace{-1.5em}
	\end{center}
\end{figure}

The temporal evolution of the relative distance $d(t)$ and the center position $R(t)$ of the vortex dipole, for the same setting as in \Fig{VortexTrajectories} ($\tilde{\mu}=6$, $d_0=150$), is shown in \Fig{RtAnddt}. 
We compare the evolutions obtained from the holographic model and from the DGPE with the trajectory \eq{doftRoft} as obtained from the HVI equations using $w_1 =1$, $d_0 = 133.5$, $R(0)= 247.2$ and the corresponding friction coefficients $C$ and $C^\prime$ stated in \Tab{DGPE_params}.
Note that, while the numerical evolutions start at $d_0 = 150$, $R(0) = 256$, we exclude from the comparison with the HVI solution the initial, outward-bended part of the holographic and DGPE trajectories which is a consequence of the initially imprinted phase structure not immediately being a solution of the full equations, see \App{AlgoVortPos}.
We find good agreement of the HVI curves with the results obtained form the DGPE and the holographic model up to times $t \simeq 600$.
Deviations caused by the finite width of the vortex cores arise for $t \gtrsim 600$.

Extrapolating the HVI trajectories according to Eqs.~\eq{doftRoft}, the intersection of the vortex paths would occur at time $t_{I}=d_{0}^{2}/8C'$, enclosing an angle $\beta$ given by
\begin{equation}
         \tan(\beta/2)
	= \frac{w_{1}d_{0}/2}{R(t_{I}) - R(0)}
	= \frac{C'}{1-C} 
	= -\frac{\delta}{\alpha}
	= \gamma\,|\ln\xi\,|\,.
\end{equation}
\Fig{VortexTrajectories} indicates these straight trajectories with parameters obtained from a DGPE simulation on a numerical grid with rescaled grid spacing  $s'=14.5$, corresponding to a lattice-unit healing length $\xi'=\xi (s/s')\equiv \xi / \hat{s} =  0.03$, for the solution with $\gamma=0.313$ and $\tau=5.76$ which was obtained by matching the holography solution on the original grid with $s=1/8$, $\xi=3.5$ as described in \App{Vortices}.
For simplification, we use the notation $\hat{s} = s'/s$ as the scale parameter in the main text.
To match the angle of the trajectories of the vortex dipole for $\tilde{\mu} = \{4.5,7.5,9\}$, we found the scale parameter to be $\hat{s} = \{114,116,119\}$.
We emphasize again that the HVI equations have been shown to describe the DGPE point vortex trajectories in the asymptotic limit $\xi\to0$.
Hence, the slightly different scalings with $\hat s$ are likely due to our matching of the HVI solutions for non-vanishing vortex core sizes. 
We list the resulting friction coefficients $C$ and $C'$ obtained from Eqs.~\eq{DragCoeffDGPE} in \Tab{DGPE_params}.

\subsection{Comparison with measurements of vortex diffusivity} 
\label{app:Diffusivity}
%
The dissipative motion of vortices has been studied experimentally and compared with the description in terms of the HVI equations \eq{HVIfull}, \eq{HVIsf} so far mainly in superfluid helium, in two-dimensional films \cite{Kim1984a.PhysRevLett.52.53,Adams1987a.PhysRevB.35.4633,Agnolet1989a.PhysRevB.39.8934,Finoretto1990a.PhysRevB.41.10994,Oda2009a.JLTP.158.262,Harris2016a.NatPhys8.788,Sachkou2019a.Science366.1480} as well as in 3D superfluids \cite{Golov2009a.JLTP.156.51,Kozik2008a.PhysRevLett.100.195302}.
Experiments with atomic Bose-Einstein condensates, however, also provide versatile platforms for the study of dissipative vortex dynamics \cite{Rosenbusch2002a.PhysRevLett.89.200403,AboShaeer2002.PhysRevLett.88.070409,Hadzibabic2008a.NJP.10.045006,Hadzibabic2009a,Henn2009a.PhysRevLett.103.045301,Neely2012a.PhysRevLett.111.235301,Kwon2014a.PhysRevA.90.063627,Moon2015a.PhysRevA.92.051601,Johnstone2019a.Science.364.1267,Gauthier2019a.Science.364.1264}.
A number of theoretical studies has been focused on the effects of dissipation on vortices in such systems, \cf{}, e.g., \cite{Fedichev1999a.PhysRevA.60.R1779,Duine2004.PhysRevA.69.053623,Kobayashi2006a.PhysRevLett.97.145301,Berloff2007a.PhysRevLett.99.145301,Jackson2009a.PhysRevA.79.053615}.

A parameter of particular interest is the friction coefficient $C'$ in the HVI equations which quantifies the motion of vortices perpendicular to the fluid flow and arises from the drag force created through interactions with the thermal excitations. 
$C'$ is related to the appearance of $\gamma\not=0$ in the DGPE, see \App{HVI}.
$C'\sim\gamma$ is usually being expressed in terms of the vortex diffusivity $D$, in two spatial dimensions defined via 
\begin{align}
    C' = \frac{2\pi\hbar\, \rho_\mathrm{s} \sigma  }{M\, k_B T}\,D\,,
\label{eq:Diffusity}
\end{align}
where $\rho_\mathrm{s}$ is the 3D superfluid {\sl mass} density, $\sigma$ the thickness of the film, $M$ the particle mass, and $T$ the temperature. 

On dimensional grounds the diffusivity has been argued to be of order $D\sim\hbar/M$ near the critical temperature of the BKT transition \cite{Ambegaokar1978a.PhysRevLett.40.783,Ambegaokar1980a}.
At temperatures in the Millikelvin to Kelvin regime, up to the BKT temperature, experiments with helium films have confirmed this value to represent, by order of magnitude, an upper limit, while $D$ has proven to be strongly temperature dependent in general, giving values of $D\sim10^{-2}\dots1\,\hbar/M$, corresponding to $C'\sim\gamma\sim10^{-2}\dots1$ \cite{Kim1984a.PhysRevLett.52.53,Adams1987a.PhysRevB.35.4633,Agnolet1989a.PhysRevB.39.8934,Finoretto1990a.PhysRevB.41.10994}.
This is considerably larger than friction coefficients on the order of $C'\sim\gamma\sim10^{-4}$ observed in atomic Bose condensates at $T\sim30\,\text{nK}$ \cite{Gauthier2019a.Science.364.1264}.
Also in recent experiments with superfluid helium films on silicon chips, coherent dynamics has been found to dominate, with very small friction $C'\sim2\dots3\times10^{-6}$ prevailing \cite{Sachkou2019a.Science366.1480}.

Note, however, that in a recent experiment with a dilute $^{23}$Na gas, strong thermal friction, $C'\sim0.01\dots0.03$, was measured at an order of magnitude higher temperature \cite{Moon2015a.PhysRevA.92.051601}, with its nearly linear temperature dependence being consistent with the prediction \cite{Fedichev1999a.PhysRevA.60.R1779}
\begin{align}
    C' \simeq 0.1\frac{4\pi\, aM^{1/2}}{\hbar\,\mu^{1/2}} k_B T\,,
\label{eq:CprimedFedichev}
\end{align}
which is based on a high-temperature ($k_B T\gg\mu$) evaluation of the analysis of \cite{Iordanskii1966JETP...22..160I}.
Moreover, the data of \cite{Moon2015a.PhysRevA.92.051601} is also matched, at the higher temperatures, by a numerical evaluation of the Zaremba-Nikuni-Griffin kinetic approach \cite{Jackson2009a.PhysRevA.79.053615} in which a generalized GPE is coupled to a Boltzmann equation for the thermal component \cite{Zaremba1999a.JLTP116.277,Allen2013a}.

Our matching procedure yields friction coefficients in the range $C'\simeq0.03\dots0.1$, \cf{}~\Tab{DGPE_params}, and thus similar to those observed in strongly dissipative systems, namely in He-films at temperatures $T \sim 1\,\text{K}$ \cite{Kim1984a.PhysRevLett.52.53,Adams1987a.PhysRevB.35.4633,Agnolet1989a.PhysRevB.39.8934,Finoretto1990a.PhysRevB.41.10994} and in thermally excited atomic condensates \cite{Moon2015a.PhysRevA.92.051601}. 

\section{Matching vortex solutions of the holographic and DGPE models} \label{app:Vortices}

\subsection{Matching the holographic and DGPE vortex profiles} \label{app:MatchingProfiles}
To compare and match the shape of the DGPE vortices to that of the holographic vortices we have considered a symmetric vortex lattice in accordance with the periodic boundary conditions in our systems.
The configuration consists of two vortices and two antivortices on a square grid evenly spaced around the center of the two-dimensional grid at $(x,y) = (256,256)$. 
In doing so we ensured a minimum mutual influence onto the individual shape of each vortex, \ie, their shape is closest to that of an individual and isolated vortex in an infinite plane.
We stress that this configuration is only taken to compare and match the spatial density profiles of the vortices. 
We then used the parameters obtained from this matching procedure as input for the dynamical simulations. 

In the holographic system, each of the vortices was imprinted at initial time $t_0$ by imposing the complex scalar field $\Phi(t, x, y, z)=|\Phi(t, x, y, z)|\exp[i\vartheta(t,x,y,z)]$ to vanish at the respective position $(x,y)=(x_{i},y_{i})$ of the $i$th vortex core along all $z$, $|\Phi(t_{0}, x_{i}, y_{i}, z)|=0$.
Moreover, the phase of $\Phi(t, x, y, z)$, was chosen to wind around the core by $2 \pi w$, $w=\pm1$, as $\vartheta_{i}(t_{0}, x, y, z)=w_{i} \mathrm{arg}[x-x_{i}+i(y-y_{i})]$, extending across the entire $(x,y)$-grid for every slice in the holographic direction \cite{Keranen2010a}.  
The phase $\vartheta$ of the field $\Phi$  is the sum of the phases around each vortex, $\vartheta=\sum_{i}\vartheta_{i}$.
It then takes roughly five unit timesteps for the vortices to build up their density profile \cite{Chesler2013a.Science341.368,Ewerz:2014tua}.

In the DGPE simulation, the vortices were prepared in the same way as far as the phase field is concerned.
For the density profile we inserted, for each vortex, the approximate analytic solution of the non-dissipative ($\gamma = 0$) Gross-Pitaevskii vortex \cite{Schakel2008a,PethickSmith}
\begin{equation}
  \label{eq:DensProfileVortex}
  |u(r)|^{2} = \frac{r^2}{2 \xi^2  + r^2}\,.
\end{equation}
Subsequently, we performed imaginary-time propagation with $\gamma=1$ for two unit timesteps $t_{\mathrm{imag}} = 2/\tau$, to relax the vortex configuration.
We stress that \Eq{DensProfileVortex} was only chosen as an approximate profile before the imaginary-time propagation causes the fields to relax to the actual vortex profiles.
Nonetheless, the functional form \eq{DensProfileVortex} in particular matches precisely the actual vortex core profile to order $\mathcal{O}(r^{2})$ around the center. 
As a result, the healing length $\xi$ determines the width of the core which thus scales as $\xi\sim\mu^{-1/2}$ with the DGPE chemical potential.

The quantitative matching of the vortex profiles in holography and DGPE requires adjusting the DGPE healing length $\xi$ such that the vortex sizes agree in both systems.
In \Fig{VortexShape} we show, for an exemplary choice 
$\tilde{T}/\tilde{T}_\text{c}=0.68$ (corresponding to $\tilde\mu=6$), the density profile of the holographic vortex (blue dots), as well as that of the DGPE vortex (orange diamonds) obtained by tuning $\xi$.
In the DGPE model, the vortex size and shape are independent of the damping parameter $\gamma$. 
For the matching of the sizes we therefore have the freedom to choose $\gamma=0$.
In practice we then tuned the DGPE healing length $\xi$ such that the density depletions of the vortices were resolved by the same number of grid points at $95\%$ of the background density.
The comparison in \Fig{VortexShape} shows that after matching the sizes, also the entire spatial profiles of the vortices are found to agree within 2 \% of the background density.
This holds true for all $\tilde{T}/\tilde{T}_\text{c}$ of the holographic system considered.
We stress that only the $\gamma$-independence of the vortex size and shape in the DGPE allows us to extract the healing length $\xi$ at $\gamma=0$ and then subsequently use this value to match the vortex dynamics to find $\gamma$ and $\tau$.

\begin{figure}
	\begin{center}
		\includegraphics[width=0.5\textwidth]{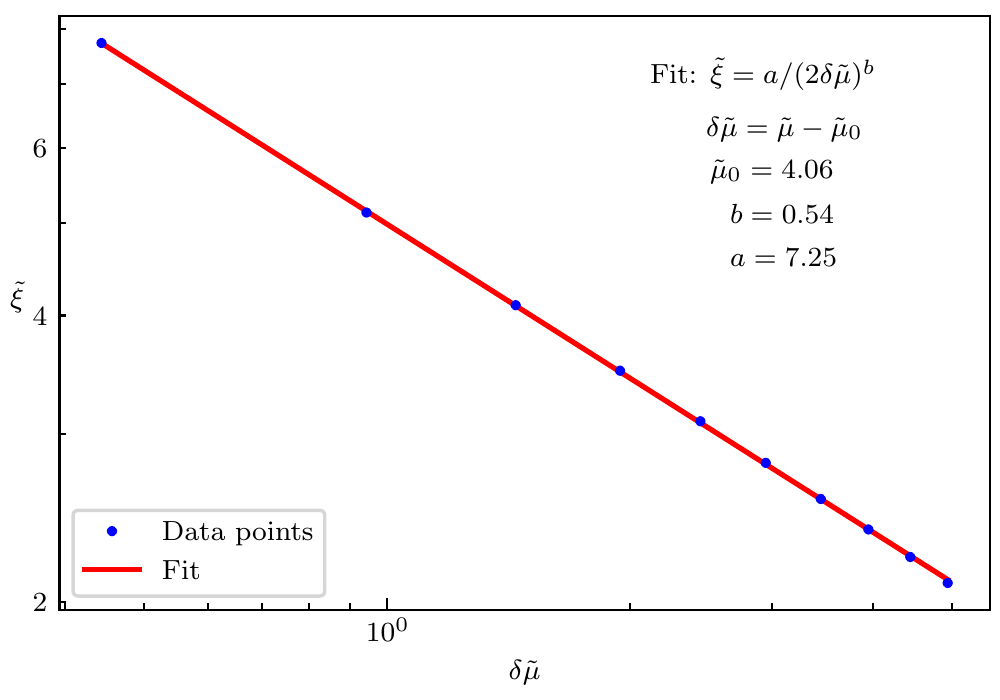}
		\caption{%
			Width $\tilde\xi$ of the holographic vortex core as a function of the holographic chemical potential $\tilde\mu=\mathbf{A}_{t}(z=0)$.
			The double logarithmic scale demonstrates that the width depends on $\delta\tilde\mu=\tilde\mu-\tilde\mu_0$ as $\tilde\xi\sim\delta\tilde\mu^{-b}$, with scaling exponent $b=0.54$.
			The fitted value $\tilde\mu_{0}=4.06$ is close to the critical value $\tilde\mu_\mathrm{c}\simeq4.064$ at which the order parameter $\psi$ vanishes.
			This behavior coincides with the dependence $\xi\sim\mu^{-1/2}$ of the DGPE healing length on the nonrelativistic chemical potential $\mu$. 
			\label{fig:MuDep}
		}
		\vspace{-1.5em}
	\end{center}
\end{figure}

Also for the holographic system the width of a vortex depends on the holographic chemical potential $\tilde{\mu}=\mathbf{A}_{t}(z=0)$.
We quantitatively studied this dependence of the vortex size on $\tilde{\mu}$ within a range of $4.5 \leq \tilde{\mu} \leq 9$ for which we found the numerical evaluation of the holographic equations feasible.
Instead of extracting the healing length for each vortex by matching a GP vortex to the holographic one, we extracted the width of the core from a Gaussian fit to the inverted core profile, $1-|u(r)|^{2}=\exp\{-r^{2}/2\tilde\xi^{2}\}$, which at small $r\lesssim\tilde\xi$ matches with the GP profile to order $\mathcal{O}(r^{2})$.
The dependence of the extracted width $\tilde\xi$ on the shifted chemical potential $\delta\tilde\mu=\tilde\mu-\tilde\mu_{0}$ is shown in \Fig{MuDep}.
The double logarithmic scale demonstrates that the width scales as $\tilde\xi\sim\delta\tilde\mu^{-b}$, with a fitted shift $\tilde\mu_{0}=4.06$ and exponent $b=0.54$. 
We note that such a behavior is expected in the superfluid regime, in the vicinity of the critical point, where  $\delta\tilde\mu$ plays the role of the gap scale in the Ginzburg-Landau theory governing the boundary dynamics of the holographic superfluid \cite{Maeda2008a.PhysRevD.78.106006}.
We furthermore found that the extracted shift $\tilde\mu_{0}$ is close to the critical chemical potential $\tilde\mu_\mathrm{c}\simeq4.064$ at which the order parameter $\psi$, \cf{}~\Eq{BoundaryExpansion}, vanishes \cite{Herzog:2008he}. 
We emphasize that the scaling coincides with the dependence $\xi\sim\mu^{-1/2}$ of the DGPE healing length on the nonrelativistic chemical potential $\mu$ which corroborates the holographic system to describe, below the critical point, a nonrelativistic superfluid \cite{Hartnoll:2008vx,Herzog:2008he,Sonner2010a.PhysRevD.82.026001,Anninos:2010sq}.
Note, however, that the holographic superfluid has an intrinsic invariance under Poincar\'e transformations of the boundary but for the vortex dynamics studied here is dominated by small velocities of the fluid and of the vortices. 

\subsection{Matching the vortex positions in the time-evolving system} \label{app:AlgoVortPos}

Once the GP healing length $\xi$ had been determined for a given $\tilde\mu$ in the holographic system, we compared and matched to each other the time evolutions of the vortices in the respective simulations.
For this we imprinted, in each system, a vortex dipole consisting of two vortices with winding numbers $w_{1,2}=\pm1$, at positions $(x_{1},y_{1})=(181,256)$, $(x_{2},y_{2})=(331,256)$ of our square grid.
We used the same procedures of presetting phase and density and a short (imaginary-time for DGPE) evolution as described in \App{MatchingProfiles}. 

To match the time-evolving vortex positions, we needed to determine the positions of the moving vortex cores with high accuracy, \ie, with a sub-grid-spacing resolution.
To be specific, we extracted the vortex locations for each unit timestep. 
For this we employed a combination of two vortex-tracking routines. 
At large vortex-antivortex separations we used a recently developed fitting algorithm \cite{Ewerz:2020}, which is based on a linear combination of two Gaussian fits for simultaneously extracting the positions of both vortices on the ($x,y$)-grid.
For small separations, on the other hand, as soon as the vortices deformed and the dipole system started to merge, we used a Newton-Raphson (NR) method (see \eg\ \cite{GALANTAI200025}) on the two-dimensional grid. 

For the fitting routine we chose a subregion of the ($x,y$)-grid that contains both vortices. 
For all results presented here, the subregion was chosen to be rectangular, with its size given by the rectangle spanned by the vortex dipole plus $40$ grid points in the positive and negative $x$ and $y$ directions.
We explicitly checked the independence of our results on the chosen subregion, given that it was sufficiently large to capture both vortices.  
In the chosen subregion we applied the combination of two 2D Gaussian fits according to
\begin{equation}
\rho(x,y) = A - A_1 \exp \left \{- \frac{\left (x - x_1 \right)^2}{2 \sigma_{x,1}^2} - \frac{\left (y - y_1\right)^2}{2 \sigma_{y,1}^2}\right \}  - A_2 \exp \left \{- \frac{\left (x - x_2 \right)^2}{2 \sigma_{x,2}^2} - \frac{\left (y - y_2\right)^2}{2 \sigma_{y,2}^2}\right \} 
\end{equation}
to determine the core positions $(x_i,y_i)$ of the vortices $i=1,2$ forming the dipole.
Here, $\sigma_{x,i}, \sigma_{y,i}$ ($i=1,2$) denote the widths of the Gaussians in $x$ and $y$ direction. 
The amplitudes $A$, $A_1$ and $A_2$ are additional fit parameters to embed the fit of the vortex dipole into the condensate background.
This enabled us to capture the entire vortex dipole in a single fit. 

The NR method is well established in the literature as an algorithm for vortex tracking in two- and three-dimensional systems, \cf\ \cite{PhysRevE.86.055301,PhysRevE.78.026601,Villois_2016}, and we therefore only briefly summarize the important steps here for the readers' convenience.
The method is based on an iterative procedure, making use of the fact that vortices are zeros in the superfluid density $\rho(x,y)$, and thus $\psi(\bm{r_i})=0$ for the position $\bm{r}_i=(x_i, y_i)$ of the $i$th vortex at a fixed timestep. 
Starting from an initial guess for the vortex position $\bm{r}_i^g$, the next better approximation is obtained from a Taylor expansion of $\psi(x,y)$ about $\bm{r}_i^g$,
\begin{align}\label{eq:NRTaylor}
  0=\psi(\bm{r}_i)=\psi(\bm{r}_i^g) + J(\bm{r}_i^g)(\bm{r}_i - \bm{r}_i^g) + \mathcal{O}\left[(\bm{r}_i - \bm{r}_i^g)^2\right]\,,
\end{align}
where $J(\bm{r}_i^g)$ denotes the Jacobian matrix of the wave function $\psi(x,y)$, 
\begin{align}
	J(\bm{r})=\begin{pmatrix}
	 \Re[ \partial_x \psi(\bm{r})]   & \Re[ \partial_y \psi(\bm{r})] \\
	\Im[ \partial_x \psi(\bm{r})] &\Im[ \partial_y \psi(\bm{r})] \\
	\end{pmatrix} \,.
\end{align}
If the Jacobian has full rank one can invert \Eq{NRTaylor} to obtain $\bm{r}_i$,
\begin{align}
	\bm{r}_i = \bm{r}_i^g - J^{-1}(\bm{r}_i^g)\,\psi(\bm{r}_i^g) + \mathcal{O}\left[(\bm{r}_i - \bm{r}_i^g)^2\right]\,.
\end{align}
Iterating this procedure, $\bm{r}_i$ converges to the desired vortex position with sub-plaquette precision. 
To evaluate the Jacobian at positions in between
full grid points, we Fourier-interpolated the field on the entire grid.
As an initial guess for the vortex position we used the approximate position obtained from locating the vortex phase winding.
We eventually stopped the iterative procedure as soon as $\lvert \psi(\bm{r_i}) \rvert  < 10^{-10} \sqrt{\rho_0}$. 

The tracking procedures allowed us to determine the position of the vortices in a precise and quasi-continuous (at unit timesteps) manner, as compared with the standard plaquette techniques, where the position can only be measured with an uncertainty of one grid point. 
For vortex-antivortex separations larger than $25$ grid points, the two tracking methods yield the same result for the vortex positions within an absolute error of $\pm 0.02$ grid points. 
This coincides with the estimated error of  the Gaussian fitting routine, \cf\ \cite{Ewerz:2020}. 
For separations smaller than $25$ grid points, the Gaussian fitting routine breaks down due the strong vortex deformations, while the NR method still tracks the vortices accurately.
Numerically, the Gaussian fitting routine is distinctly faster then the NR method. 
We therefore combined the two methods, using Gaussian fits in the regime $d(t)>25$ while the NR method was applied for all smaller separations. 

Being able to determine accurately the continuous trajectories of the vortex dipole obtained in holography and from the DGPE, we can match them in space and time by tuning the DGPE parameters $\gamma$ and $\tau$, respectively.
We found that, for a fixed healing length, \ie, a particular width of the vortices, the position-space trajectory of the vortex dipole in the DGPE simulation is solely determined by the damping parameter $\gamma$, which provides a quantitative  measure of the dissipation. 
For an agreement of the dynamics on the temporal grids of the holographic and DGPE simulations we adjusted the DGPE scale parameter $\tau$, which is defined in \App{dlDGPE}.

Our findings show that the damping parameter $\gamma$ has only a small dependence on the temperature ratio $\tilde{T}/\tilde{T}_\text{c}$ of the holographic superfluid, see \Tab{DGPE_params}. 
Hence, the dissipation of the holographic system is only mildly dependent on its temperature, at least in the probe limit of the holographic superfluid that we consider here. 
The parameter $\tau$ which controls the relative scaling of temporal and spatial lattice units, on the other hand, varies substantially with $\tilde{T}/\tilde{T}_\text{c}$, see  \Tab{DGPE_params}.

Finally, we want to comment on a peculiar feature that the numerically calculated trajectories of the vortex dipole exhibit at early times of the evolution, both in holography and DGPE. 
As can be seen in \Fig{VortexTrajectories}, the dynamics causes the trajectories of the vortices to initially bend slightly outwards instead of following straight lines as would be expected from the HVI equations, see \App{HVI}. 
We note that the effect is small and becomes clearly visibly only due to applying a tracking algorithm with sub-plaquette resolution. 
The effect originates from the initial phase imprinting of the vortex dipole. 
The imprinting of two separate phase fields for the vortices by simply adding their phases at all points across the lattice is in fact incompatible with a true two-vortex solution of the equations of motion. 
The deviation of the phase field from an actual solution can be shown to lead to an outward velocity of the superfluid that acts on the vortices accordingly \cite{Ewerz:2020}. 
Similar effects have also been observed for soliton solutions in GP simulations \cite{Christov:2018ecz,Christov:2018wsa}. 
During the numerical evolution of the vortex pair, the phase adjusts itself to a full two-vortex solution, after which the vortices follow straight lines, see \Fig{VortexTrajectories}. 
In the example shown in there, the phase healing process terminates at time $t \simeq 190$ (corresponding to a vortex separation of $d_0=133.5$), and we use only the subsequent evolution for the determination of the friction coefficients of the HVI equations, see also \App{HVI}. 

\end{appendix}
%

\end{document}